\documentclass[twocolumn,showpacs,preprintnumbers,amsmath,amssymb,superscriptaddress,nofootinbib,english]{revtex4-1}
\usepackage{times,amsmath,amsfonts,amssymb,epstopdf}
\usepackage{graphicx}
\usepackage{dcolumn}
\usepackage{bm}
\usepackage{epsfig}
\usepackage{graphicx}
\usepackage{hyperref}
\usepackage[usenames]{color}
\usepackage{url}
\usepackage[normalem]{ulem}
\usepackage[T1]{fontenc}
\usepackage[dvipsnames]{xcolor}

\def\be{\begin{equation}}
\def\ee{\end{equation}}
\def\ba{\begin{eqnarray}}
\def\ea{\end{eqnarray}}
\begin{document}

\title{Linear perturbations in K-mouflage cosmologies with massive neutrinos}

\author{Alexandre Barreira}\email{Email: a.m.r.barreira@durham.ac.uk}
\affiliation{Institute for Computational Cosmology, Department of Physics, Durham University, Durham DH1 3LE, UK}
\affiliation{Institute for Particle Physics Phenomenology, Department of Physics, Durham University, Durham DH1 3LE, UK}
\author{Philippe Brax}\email{Email: philippe.brax@cea.fr}
\affiliation{Institut de Physique Theorique, CEA, IPhT, CNRS, URA 2306, F-91191Gif/Yvette Cedex, France}
\author{Sebastien Clesse}\email{Email: sebastien.clesse@unamur.be}
\affiliation{Namur Center of Complex Systems (naXys), Department of Mathematics, University of Namur, Rempart de la Vierge 8, 5000 Namur, Belgium}
\author{Baojiu Li}\email{Email: baojiu.li@durham.ac.uk}
\affiliation{Institute for Computational Cosmology, Department of Physics, Durham University, Durham DH1 3LE, UK}
\author{Patrick Valageas}\email{Email: patrick.valageas@cea.fr ; Author names listed by alphabetical order. }
\affiliation{Institut de Physique Theorique, CEA, IPhT, CNRS, URA 2306, F-91191Gif/Yvette Cedex, France}


\begin{abstract}

We present a comprehensive derivation of linear perturbation equations for different matter species, including photons, baryons, cold dark matter, scalar fields, massless and massive neutrinos, in the presence of a generic conformal coupling. Starting from the Lagrangians, we show how the conformal transformation affects the dynamics. In particular, we discuss how to incorporate consistently the scalar coupling in the equations of the Boltzmann hierarchy for massive neutrinos and the subsequent fluid approximations. We use the recently proposed K-mouflage model as an example to demonstrate the numerical implementation of our linear perturbation equations. K-mouflage is a new mechanism to suppress the fifth force between matter particles induced by the scalar coupling, but in the linear regime the fifth force is unsuppressed and can change the clustering of different matter species in different ways. We show how the CMB, lensing potential and matter power spectra are {affected} by the fifth force, and find {ranges of K-mouflage parameters whose effects could be seen observationally.}
{We also find that the scalar coupling can have the nontrivial effect of shifting the amplitude  of the power spectra of the lensing potential and density fluctuations in opposite directions, although both probe the overall clustering of matter.} This paper can serve as a reference for those who work on generic coupled scalar field cosmology, or those who are interested in the cosmological behaviour of the K-mouflage model.

\end{abstract}


\maketitle

\section{Introduction}

The confirmation that our Universe is experiencing a phase of accelerated expansion \cite[see, e.g.,][]{sne_data,cmb_data,bao_data} has provoked extensive research aiming to find out an underlying driving force. The majority of models proposed so far involve one or more scalar fields, which experience self interactions either through a self potential, such as the quintessence model \cite[e.g,][]{quintessence1,quintessence2}, or via non-standard kinetic terms, such as the K-essence model \cite[e.g.,][]{k-essence1,k-essence2}. If a scalar field is present, it is both theoretically and phenomenologically interesting to assume that it interacts with either matter or curvature, considerations of which have led to the developments of coupled quintessence \cite[e.g.,][]{c-quintessence} and extended quintessence \cite[e.g.][]{e-quintessence} models, with both types of models having a standard kinetic term for the scalar field.

The existence of a scalar field coupling to matter or curvature can be problematic, because the scalar field can mediate a so-called fifth force between matter particles, in conflict with local gravity tests \cite[e.g.,][]{will2014}. To avoid this problem, it is often assumed that either the scalar field does not interact with baryonic components of matter, such as in the coupled dark energy model, or there is some mechanism to suppress the fifth force where gravity experiments are carried out. The latter idea may sound odd, but it can be a natural consequence of the nonlinearity of the self-interacting potential of the scalar field. Some well known examples of such `screening mechanisms' are the chameleon \cite{kw2004}, dilaton \cite{bbds2010} and symmetron \cite{hk2010} mechanisms. In these models, the interaction of matter can give a heavy mass to the scalar field \cite{kw2004}, or trap it to values that make the interaction strength very weak \cite{bbds2010,hk2010}, in regions of high matter density. In these models the kinetic term of the scalar field is assumed to be standard.

Non-standard (non-canonical) kinetic terms can also naturally lead to suppression of the fifth force, such as in the case of the Dvali-Gabadadze-Poratti (DGP) \cite{dgp} and the Galileon \cite{galileon1,galileon2} models, where the matter density, or equivalently $\nabla^2\Phi$, is high. This is known as the Vainshtein mechanism \cite{vainshtein}. Another example of a coupled scalar field with a non-standard kinetic term is the K-mouflage model \cite{bv2014a,bv2014b}, which is a K-essence-type scalar field coupled to matter.

The idea of K-mouflage offers a novel perspective on the screening of scalar interactions in dense environments. It differs from the chameleon mechanism, for which the screening takes place in regions where the Newtonian potential is larger than a threshold value determined by the scalar field itself. It is also different from the Vainshtein mechanism that operates in Galileon models, in which the screening occurs in regions of large scalar curvature. Instead, in the case of K-mouflage, the screening happens in regions where the {\it gravitational acceleration} is large enough. The phenomenology of the K-mouflage screening can therefore be qualitatively different from that of the chameleon and Vainshtein screenings, and has been studied less intensively so far (see, e.g., Sec.~II of \cite{bv2014a} for a brief comparison of these three types of screening mechanisms).

In the static regime, the existence of a K-mouflage radius, below which the screening happens, and  of a static solution of the Klein-Gordon equation, depends crucially on the form of the Lagrangian, $M^4 K(\sigma)$, where $\sigma\equiv(\nabla\varphi)^2/2M^4$, $M$ is the dark energy scale and $K(\sigma)$ is a nonlinear function {(cf.~Eq.~(\ref{eq:scalar_lagrangian}) below)}; if $K(\sigma)=\sigma$, {the kinetic term becomes canonical}. K-mouflage models can also be extended to non-static cosmological backgrounds for a restricted class of $K$-functions.  {Healthy K-mouflage models are those where the screening can be achieved in the static regime and cosmological solutions can be defined down to arbitrarily early cosmic times.  This implies that the potentials defined as $W_\pm (y)\equiv y K'\left(\pm y^2 /2\right)$ are monotonic and go to infinity at large positive $y$\footnote{{Here, $y=\sqrt{\pm 2 \sigma}$} and note that $\sigma\propto-(\vec{\partial}\varphi)^2/2<0$, where $\vec{\partial}$ denotes the spatial derivative, in the static case.}. Moreover, the {value of} $K'\left(-y^2/2\right)$ must be large for large enough $y^2$ to suppress the scalar fifth force inside the K-mouflage radius -- the $\prime$ above means a derivative w.r.t. to argument of $K$ (see~\cite{bv2014c} for more details).}

Cosmologically, the effects of the scalar interaction appear both at the background {and perturbation levels}. At  the background level~\cite{bv2014a}, healthy K-mouflage models all cross the phantom divide in the recent past and the effective energy density of the scalar becomes negative in the distant past. This does not lead to instabilities as the Hubble rate squared is always positive: the K-mouflage field is subdominant, i.e., cosmologically screened, in dense cosmological densities. At late times, the growth of density perturbations is changed as the effective gravitational strength can either be increased ($K'>0$) or decreased ($K'<0$) in a scale independent way \cite{bv2014b}.
{Examples of healthy K-mouflage models are polynomials whose higher degree monomial, $K_0 \sigma^m$, is such that $K_0 > 0$ and $m$ is an odd integer, where $K_0$ and $m$ are model parameters, see Eq.~(\ref{K-def-K0-m}). Models with $K_0 < 0$ have a ghost-like behaviour and require a contrived UV cutoff at a rather low energy scale. In this paper, we will focus on cases with either $m = 2$ or $m = 3$, and $K_0$ of both signs. The reader should bear in mind that this is done for illustration purposes. Only the case with $m = 3$ and $K_0 > 0$ is both healthy and ghost-free (in both the cosmological regime and the small-scale static regime).}



In this paper, we numerically study the evolution of linear perturbations in the K-mouflage model. One of our main goals is to analyse the model predictions for observables such as the CMB temperature, CMB lensing, and matter power spectra.

We shall start by deriving the perturbation equations in the presence of a conformally coupled scalar field. Although some of these equations have been derived in the past and are scattered in the literature, we feel that a more complete and consistent derivation is needed, for the following reasons:
\newline
\indent(i) In cosmological studies, we are often interested in a universal coupling of the scalar field with all matter species, and thus the effect of the coupled scalar field must be consistently included for all these species. We shall do this from the  Lagrangian level.
\newline
\indent(ii) some matter species, e.g., massive neutrinos, have not been {extensively} studied in the presence of a scalar coupling, although the role played by massive neutrinos in cosmology is increasingly becoming a topic of interest. There are previous works along this direction, \cite[e.g.,][]{brookfield2006,ik2008}, but there the neutrino perturbation equations are derived in the synchronous gauge rather than in a more general gauge-invariant formalism, and these works are focused on a coupling between the scalar field and massive neutrinos only. A subtler point relates to the neutrino equations in the so-called fluid approximation, which are not present in those works -- this is not necessarily problematic, but we should bear in mind that standard Boltzmann codes, such as the one used in this paper and in Ref.~\cite{brookfield2006}, usually silently switch to this approximation at late times for efficiency considerations, and inconsistency would arise if these approximation equations are not modified accordingly to take into account the scalar field coupling. Here we will present the modified equations in the fluid approximation for neutrinos.

To obtain cosmological predictions, we have modified the {\sc Camb} code~\cite{camb} to solve our linear perturbation equations. In this paper, it is not our goal to perform a thorough exploration of the parameter space of the K-mouflage model. Instead, we shall focus on a number of illustrative parameter values to try to build intuition about the regions of the parameter space that are more likely to be ruled out, or alternatively, provide a good fit to the data. We shall pay particular attention to the potential degeneracies between the K-mouflage parameters and the mass of active neutrinos.

The present paper is organised as follows. In \S~\ref{sect:equations} we will describe the conformal transformation between the Jordan and Einstein frames, and apply this to the Lagrangian densities of photons, neutrinos (massless and massive), classical particles (baryonic and cold dark matter), and general scalar fields to derive their respective conservation equations in the Einstein frame, where our calculations are done. The scalar field is a K-mouflage field for this work, although some of our derivations hold generically for any coupled scalar field. In \S~\ref{sect:pert_equations} we present the covariant and gauge invariant linear perturbation equations for standard gravity and, using the results of \S~\ref{sect:equations}, derive the perturbation equations for matter species, with particular attention paid to the case of massive neutrinos. In Sec.\ref{sec:results} we present and discuss our numerical results. We start by describing the details of our numerical setup and then discuss the model predictions for the CMB temperature, CMB lensing and matter power spectra. Finally, we summarise our findings in Sec.~\ref{sect:conclusions}, where we also briefly compare the K-mouflage model with other popular modified gravity models.

%

\section{Equations in conformally coupled scalar field cosmology}

\label{sect:equations}

\subsection{The general field equations}

\label{subsect:general_eqns}

The Einstein Hilbert action is
\begin{eqnarray}\label{eq:action}
S &=& \int{\rm d}^4x \sqrt{-g}\left[\frac{1}{2}M^2_{\rm Pl}R + \mathcal{L}_\varphi(\varphi)\right]+S_{\rm m},
\end{eqnarray}
{with
\begin{eqnarray}\label{eq:S_m}
S_{\rm m} &=&  \sum_{i}\int{\rm d}^4x \sqrt{-\tilde{g}}\mathcal{\tilde{L}}_{\rm m}\left(\tilde \psi_{\rm m}^{(i)}, \tilde{g}_{\mu\nu}\right),
\end{eqnarray}
where $g$ ($\tilde{g}$) is the determinant of the Einstein (Jordan) frame metric tensor $g_{\mu\nu}$ ($\tilde{g}_{\mu\nu}$), $\mathcal{\tilde{L}}_{\rm m}$ is the matter Lagrangian density in the Jordan frame} and $\tilde \psi^{(i)}_{\rm m}$ symbolically denotes the $i$th species of matter fields. The Jordan and Einstein frame metric tensors are related by a conformal transformation,
\begin{eqnarray}\label{eq:conformal_transform}
\tilde{g}_{\mu\nu} &=& A^2(\varphi)g_{\mu\nu},
\end{eqnarray}
with $A$ a function of the scalar field $\varphi$. {Above and throughout,} $M_{\rm pl}$ is the reduced Planck mass, and it is related to Newton's constant $G$ by $M^{-2}_{\rm Pl}=8\pi G$.

It can be shown straightforwardly that the Christoffel symbols in the two frames are related by
\begin{eqnarray}\label{eq:christoffel_transform}
\Gamma^\lambda_{\mu\nu} &=& \tilde{\Gamma}^{\lambda}_{\mu\nu} - \left[\delta^{\lambda}_{\mu}\left(\ln A\right)_{,\nu}+\delta^{\lambda}_{\nu}\left(\ln A\right)_{,\mu}-g_{\mu\nu}\left(\ln A\right)^{,\lambda}\right],\ \ \
\end{eqnarray}
where a comma denotes the partial derivative $\varphi_{,\mu}\equiv\partial\varphi/\partial x^\mu$, and $\varphi^{,\mu}\equiv g^{\mu\nu}\varphi_{,\nu}$.

In the Jordan frame, matter is uncoupled to the scalar field and the energy momentum tensor for a given species (the superscript $^{(i)}$ is dropped to lighten the notation) is defined as
{
\begin{eqnarray}\label{eq:T_munu_jordan}
\tilde{T}_{\mu\nu} &=& -\frac{2}{\sqrt{-\tilde{g}}}\frac{\delta\left[\sqrt{-\tilde{g}}\mathcal{\tilde{L}}_{\rm m}\left(\tilde \psi_{\rm},\tilde{g}_{\mu\nu}\right)\right]}{\delta\tilde{g}^{\mu\nu}},
\end{eqnarray} }
which satisfies the following conservation equation
\begin{eqnarray}\label{eq:conservation_jordan}
\tilde{\nabla}_\nu\tilde{T}^\nu_{\ \mu} &=& 0,
\end{eqnarray}
where $\tilde{\nabla}$ is the covariant derivative compatible with the metric $\tilde{g}_{\mu\nu}$. {The lack of a coupling between the scalar field and matter in the Jordan frame is an assumption of this paper. In practice, if all matter species are coupled to the scalar field conformally in the same way in the Einstein frame, as we assume here, one could always redefine the Jordan-frame metric to remove the coupling in the latter. Theories involving disformal couplings or different couplings for different matter species can be more complicated, and will not be covered here.}

Similarly, the energy momentum tensor defined in the Einstein frame is
\begin{eqnarray}\label{eq:T_munu_einstein}
{T}_{\mu\nu} &=& -\frac{2}{\sqrt{-{g}}}\frac{\delta\left[\sqrt{-g}\mathcal{L}_{\rm m}\left(\psi_{\rm},A(\varphi),{g}_{\mu\nu}\right)\right]}{\delta{g}^{\mu\nu}},
\end{eqnarray}
which satisfies the following (non)conservation equation
\begin{eqnarray}\label{eq:conservation_einstein}
{\nabla}_\nu{T}^\nu_{\ \mu} &=& \frac{{\rm d}\ln A(\varphi)}{{\rm d}\varphi}T\nabla_{\mu}\varphi,
\label{non}
\end{eqnarray}
where ${\nabla}$ is the covariant derivative compatible with the metric ${g}_{\mu\nu}$ and {$T = T_\mu^\mu$}.  The energy-momentum tensor $T_{\mu\nu}$ is related to $\tilde{T}_{\mu\nu}$ by\footnote{This can be done by noticing that in Eqs.~(\ref{eq:T_munu_jordan}) and (\ref{eq:T_munu_einstein}) the terms in the brackets are the same because the matter action is invariant under the conformal transformation. Then by using Eq.~(\ref{eq:conformal_transform}) it {is straightforward to show that $T_{\mu\nu}=A^2(\varphi)\tilde{T}_{\mu\nu}$. }}
\begin{eqnarray}\label{eq:emt_transform}
T^{\mu}_{\ \nu} &=& A^4(\varphi)\tilde{T}^{\mu}_{\ \nu},
\end{eqnarray}
where indices for (un)tildered quantities are raised and lowered by the (un)tildered metric. One can check Eq.~(\ref{eq:conservation_einstein}) by using Eqs.~(\ref{eq:christoffel_transform}, \ref{eq:conservation_jordan}, \ref{eq:emt_transform}).

In the next few subsections, we will look at the individual matter species and see how the above equations hold for each of them.

\subsection{Photons}

\label{subsect:photon_eqns}

In the Jordan frame, the action for photons is
\begin{eqnarray}
S_{\gamma} &=& \int{\rm d}^4x\sqrt{-\tilde{g}}\frac{1}{4\alpha}\tilde{F}^{\mu\nu}\tilde{F}_{\mu\nu},
\end{eqnarray}
where $\alpha$ is the gauge coupling constant. To change this to the Einstein frame, we define a new gauge field strength as
\begin{eqnarray}
F_{\mu\nu} &\equiv& \tilde{F}_{\mu\nu},\\
F^{\mu\nu} &\equiv& g^{\mu\alpha}g^{\nu\beta}F_{\alpha\beta}\ =\ A^4(\varphi)\tilde{g}^{\mu\alpha}\tilde{g}^{\nu\beta}\tilde{F}_{\alpha\beta}\ =\ A^4(\varphi)\tilde{F}^{\mu\nu},\nonumber
\end{eqnarray}
and the above action can be re-expressed as
\begin{eqnarray}
S_{\gamma} &=& \int{\rm d}^4x\sqrt{-{g}}\frac{1}{4\alpha}{F}^{\mu\nu}{F}_{\mu\nu},
\end{eqnarray}
leaving $\alpha$ unchanged.

From the above actions, using Eqs.~(\ref{eq:T_munu_jordan}) and (\ref{eq:T_munu_einstein}), one obtains the energy momentum tensors for photons in the two frames:
\begin{eqnarray}
\tilde{T}^{\mu}_{\ \nu} &=& \tilde{F}^{\mu\lambda}\tilde{F}_{\nu\lambda}-\frac{1}{4}\delta^{\mu}_{\ \nu}\tilde{F}^{\alpha\beta}\tilde{F}_{\alpha\beta},\\
{T}^{\mu}_{\ \nu} &=&{F}^{\mu\lambda}{F}_{\nu\lambda}-\frac{1}{4}\delta^{\mu}_{\ \nu}{F}^{\alpha\beta}{F}_{\alpha\beta},
\end{eqnarray}
so that Eq.~(\ref{eq:emt_transform}) is satisfied as expected.

In the case of photons, note that the trace $T\equiv T^{\mu}_{\ \mu}=0$ in Eq.~(\ref{eq:conservation_einstein}), so that $T^{\mu\nu}$ is conserved even in the Einstein frame.

\subsection{Neutrinos}

\label{subsect:neutrino_eqns}

Neutrinos are fermions and their action in the Jordan frame can be written as
\begin{eqnarray}
S_\nu &=& \int{\rm d}^4x\sqrt{-\tilde{g}}\left[i\bar{\tilde{\Psi}}\tilde{\gamma}^\mu\tilde{D}_{\mu}\tilde{\Psi}-\tilde{m}\bar{\tilde{\Psi}}\tilde{\Psi}\right],
\end{eqnarray}
where $\tilde{\Psi}$ denotes a Dirac fermion field, $\bar{\tilde{\Psi}}$ its conjugate, $\tilde{m}$ its mass, and $\tilde \gamma^\mu$ are the Dirac matrices satisfying
\begin{eqnarray}
\tilde{\gamma}^\mu\tilde{\gamma}^\nu+\tilde{\gamma}^\nu\tilde{\gamma}^\mu &=& 2\tilde{g}^{\mu\nu}\mathbb{I}
\end{eqnarray}
with $\mathbb{I}$ being the identity matrix, and
\begin{eqnarray}
\tilde{D}_{\mu} &{=}& \mathbb{I}\frac{\partial}{\partial x^{\mu}}+\frac{1}{4}\tilde{\omega}_{\lambda\rho\mu}\tilde{\gamma}^\lambda\tilde{\gamma}^\rho
\end{eqnarray}
is the covariant derivative of a spinor with respect to the connection $\tilde{\omega}_{\lambda\rho\mu}$. 

Transforming from the Jordan to the Einstein frame, from the above relations we have
\begin{eqnarray}
\gamma^\mu &=& A(\varphi)\tilde{\gamma}^\mu,\\
\omega_{\lambda\rho\mu}\gamma^{\lambda}\gamma^{\rho} &=& \tilde{\omega}_{\lambda\rho\mu}\tilde{\gamma}^{\lambda}\tilde{\gamma}^{\rho}-6\left[\ln A(\varphi)\right]_{,\mu}\mathbb{I}.
\end{eqnarray}
If we further consider the following definitions:
\begin{eqnarray}
\Psi &=& A^{3/2}(\varphi)\tilde{\Psi},\\
\label{eq:mass_varying} m &=& A(\varphi)\tilde{m},
\end{eqnarray}
then the above fermion action can be recast in canonical form as
\begin{eqnarray}
S_\nu &=& \int{\rm d}^4x\sqrt{-{g}}\left[i\bar{{\Psi}}{\gamma}^\mu{D}_{\mu}{\Psi}-{m}\bar{{\Psi}}{\Psi}\right].
\end{eqnarray}
Therefore, if we assume, rather reasonably, that in the Jordan frame the bare mass of the fermionic particle, $\tilde{m}$, is a constant, then in the Einstein frame the mass depends on the scalar field $\varphi$, and changes in time and space {via Eq.~(\ref{eq:mass_varying}).} {Since the field redefinitions do not affect the spinor indices of fermions, the same reasoning applies to Majorana spinors.  Majorana masses are thus also rescaled by a factor $A(\varphi)$ in the Einstein frame.
}

Let us consider now the energy momentum tensor of neutrinos on a FRW background. In the Jordan and Einstein frames, the line elements for the background universe can be written respectively as
\begin{eqnarray}
{\rm d}\tilde{s}^2 &=& \tilde{a}^2\left({\rm d}t^2-{\rm d}{\bf x}^2\right),\\
{\rm d}{s}^2 &=& {a}^2\left({\rm d}t^2-{\rm d}{\bf x}^2\right),
\end{eqnarray}
where $\tilde{a}$ and $a$ are the scale factors in these two frames, and they satisfy
\begin{eqnarray}
\tilde{a} &=& A(\varphi)a,
\end{eqnarray}
according to Eq.~(\ref{eq:conformal_transform}).

Without loss of generality, consider active neutrinos whose mass can reach up values of a few $\rm eV$ \cite{gc2009, gd2013}. At early times, before decoupling from other species, these neutrinos satisfy the equilibrium Fermi-Dirac (FD) distribution:
\begin{eqnarray}
f_0 &=& f_0(\epsilon)\ =\ \frac{g_s}{\hbar^3}\frac{1}{1+\exp\left(\epsilon/k_{\rm B}Ta\right)},
\end{eqnarray}
where $\hbar$ is the reduced Planck constant, $k_{\rm B}$ is the Boltzmann constant, $T$ is the equilibrium temperature at scale factor $a$, $g_s$ is the number of fermionic degrees of freedom, and
\begin{eqnarray}
\epsilon &\equiv& \sqrt{q^2+(ma)^2}
\end{eqnarray}
is the energy of a neutrino particle with mass $m$ and comoving momentum $q$. Note that we have not specified which frame is used in the above expression, but instead tried to make general statements (hence no tildes are used until we start talking about frames below).

Because neutrinos are highly relativistic when they decouple, then we have $\epsilon\approx q\gg ma$, and so the distribution before decoupling can be written as
\begin{eqnarray}
f_0(q) &=& \frac{g_s}{1+\exp\left(q\right)},
\end{eqnarray}
in which, and in what follows, the unit $\hbar=1$ is used and $q$ is expressed in units of $k_{\rm B}Ta$. In a unperturbed universe $Ta$ is a constant equal to the temperature today, $T_0$.

The decoupling of neutrinos could be approximately considered as an instantaneous process, in which case the equilibrium distribution above is preserved after neutrino decoupling \cite{lp2006}, since the momentum and the temperature redshift in the same way. We neglect any possible effects of a scalar coupling on the neutrino decoupling, which happens at very early times when neutrinos are highly relativistic so that the scalar field is essentially decoupled from it (though the neutrino mass could still be time varying).

In a perturbed Universe, $f$ is no longer a strict FD distribution, but instead can have time and space dependences:
\begin{eqnarray}
f({\bf x}, {\bf q}, t) &=& f_0(q)\left[1+\Psi({\bf x}, {\bf q}, t)\right]\nonumber\\
&=& f_0(q)\left[1+\Psi({\bf x}, q, {\bf n}, t)\right],
\end{eqnarray}
in which $\Psi$ (not to be confused with the fermion field above) denotes the deviation from the FD distribution. In addition to the spatial and time dependences, $f$ also depends on ${\bf q}$, in particular its direction ${\bf n}$.

The components of the energy momentum tensor are given by
\begin{eqnarray}
\label{eq:neutrino_emt_1} T^{0}_{\ 0} &=& a^{-4}\int{\rm d}\Omega{\rm d}qq^2\epsilon f_0(q)\left[1+\Psi({\bf x}, q, {\bf n}, t)\right],\\
\label{eq:neutrino_emt_2} T^0_{\ i} &=& a^{-4}\int{\rm d}\Omega{\rm d}qq^3n_i f_0(q)\Psi({\bf x}, q, {\bf n}, t),\\
\label{eq:neutrino_emt_3} T^{i}_{\ j} &=& -a^{-4}\int{\rm d}\Omega{\rm d}q\frac{q^4}{\epsilon}n_in_jf_0(q)\left[1+\Psi({\bf x}, q, {\bf n}, t)\right],\ \ \
\end{eqnarray}
in which
$n_i$ is the unit vector in the $i$th direction and ${\rm d}\Omega$ is the solid angle of the volume element in momentum space, ${\rm d}^3{\bf q}$. One important observation here is that, $\epsilon$ in these expressions depends on the combination $am=\tilde{a}\tilde{m}$, such that the integrations above are the same in both the Jordan and the Einstein frames. Consequently, Eq.~(\ref{eq:emt_transform}) is satisfied for both massive and massless neutrinos (as it should be), because $\tilde{\epsilon}=\epsilon$, and the only transformation of $T^{\mu}_{\ \nu}$ between the Einstein and Jordan frames in the above equations is through the scale factor $a$ therein.

Recall that $T^\mu_{\ \nu}$ is not conserved in the Einstein frame, even though it has the same functional form as $\tilde{T}^{\mu}_{\ \nu}$ (though with the quantities expressed in the Einstein frame). This is because in this frame the mass of the neutrinos depends explicitly on $\varphi$ (cf.~Eq.~(\ref{eq:mass_varying})). For massless neutrinos, on the other hand, due to the vanishing trace of the energy-momentum tensor, we have that $\tilde{\nabla}_\nu \tilde{T}^{\nu}_\mu = {\nabla}_\nu {T}^{\nu}_\mu = 0$.

\subsection{Dark matter and baryons}

\label{subsect:matter_eqns}

In the context of cosmological structure formation, it is reasonable to treat cold dark matter particles and baryons as free (collisionless) point masses at the microscopic level, and the Lagrangian is given by $\tilde{L}=-\Gamma\tilde{m}$, where $\Gamma\equiv{\rm d}s/{\rm d}t$ is the relativistic boost, with ${\rm d}s$ and ${\rm d}t$ being respectively the proper and physical times. In the Jordan frame, the action is given by $S_m=\int\tilde{L}{\rm d}t$ and can be re-expressed as
\begin{eqnarray}\label{eq:xxx}
S_m &=& -\int{\rm d}^4x{\tilde{m}}\sum_{i=1}^N\sqrt{\tilde{g}_{\mu\nu}\frac{{\rm d}x^\mu}{{\rm d}t}\frac{{\rm d}x^\nu}{{\rm d}t}}{\delta}^{(3)}({\bf x}-{\bf y}_i),
\end{eqnarray}
in which $\tilde{m}$ stands generally for the bare mass of the particles, and the Dirac $\delta$-function reflects the fact that the point mass is located at position ${\bf y}_i$. In this expression, we have assumed that the system contains $N$ particles for illustration purposes.


In the Einstein frame, the action structure remains the same, but must be expressed in terms of the metric $g_{\mu\nu}$ and a re-defined mass :
\begin{eqnarray}
S_m &=& -\int{\rm d}^4x{{m}}\sum_{i=1}^N\sqrt{{g}_{\mu\nu}\frac{{\rm d}x^\mu}{{\rm d}t}\frac{{\rm d}x^\nu}{{\rm d}t}}{\delta^{(3)}}({\bf x}-{\bf y}_i),\ \ \
\end{eqnarray}
where $m\equiv A(\varphi)\tilde{m}$. As in the case of massive neutrinos, in the Einstein frame the particle mass depends on the scalar field and therefore can vary in both space and time.

By applying Eqs.~(\ref{eq:T_munu_jordan}, \ref{eq:T_munu_einstein}), one finds the energy momentum tensor in the two frames as
\begin{eqnarray}
\label{eq:point_particle_emt_jordan} \tilde{T}^\mu_{\ \nu}(x) &=& \frac{\tilde{m}}{\sqrt{-\tilde{g}}}\sum_{i=1}^N\left[\tilde{g}_{\alpha\beta}\frac{{\rm d}x^\alpha}{{\rm d}t}\frac{{\rm d}x^\beta}{{\rm d}t}\right]^{-1/2}\nonumber\\
&&~~~~~~~~~~~~~~~~~~~~~\times\tilde{g}_{\nu\lambda}\frac{{\rm d}x^\mu}{{\rm d}t}\frac{{\rm d}x^\lambda}{{\rm d}t}{\delta^{(3)}}({\bf x}-{\bf y}_i),\\
\label{eq:point_particle_emt_einstein} {T}^\mu_{\ \nu}(x) &=& \frac{m}{\sqrt{-{g}}}\sum_{i=1}^N\left[{g}_{\alpha\beta}\frac{{\rm d}x^\alpha}{{\rm d}t}\frac{{\rm d}x^\beta}{{\rm d}t}\right]^{-1/2}\nonumber\\
&&~~~~~~~~~~~~~~~~~~~~~\times{g}_{\nu\lambda}\frac{{\rm d}x^\mu}{{\rm d}t}\frac{{\rm d}x^\lambda}{{\rm d}t}{\delta^{(3)}}({\bf x}-{\bf y}_i).
\end{eqnarray}
A quick inspection confirms that the above equations satisfy Eq.~(\ref{eq:emt_transform}). Note that the energy momentum tensor has mass dimension 4 as expected, as the 3D Dirac function $\delta^{(3)}({\bf x})$ has mass dimension 3.

Equations (\ref{eq:point_particle_emt_jordan}, \ref{eq:point_particle_emt_einstein}) hold for a number of discrete point particles, while in the real world cold dark matter and baryons are usually collectively treated as fluids on macroscopic scales. To this end, the standard practice is to perform a volume average in microscopically large but macroscopically small volumes, and the resulting energy momentum tensor describes an effective fluid. The relevant perturbed equations will be derived in Secs.~\ref{subsubsec:_matter_quations_cdm} and \ref{subsubsec:matter_equations_baryons}.

In the Einstein frame, although the particle number of nonrelativistic species is conserved, their energy-momentum tensor is not, because of the varying particle mass induced by the scalar coupling (cf.~Eq.~(\ref{eq:mass_varying})). As a result, it is convenient to separate the effects of the varying mass by writing the energy momentum tensor for matter as
\begin{eqnarray}\label{eq:hat_T_munu}
T^\mu_{\ \nu} &\equiv& A(\varphi)\hat{T}^{\mu}_{\ \nu},
\end{eqnarray}
so that $\hat{T}^\mu_{\ \nu}$ is conserved at the background level. This can be checked by substituting the above relation into Eq.~(\ref{eq:conservation_einstein}) to get
\begin{eqnarray}\label{eq:conservation_einstein_matter}
\nabla_\mu\hat{T}^\mu_{\ \nu} &=& -\frac{{\rm d}\ln A(\varphi)}{{\rm d}\varphi}\left(\hat{T}^\mu_{\ \nu}-\hat{T}\delta^\mu_{\ \nu}\right)\nabla_\mu\varphi,
\end{eqnarray}
where the covariant derivatives are still taken with respect to the metric $g_{\mu\nu}$ and $\hat{T}=\hat{T}^\mu_{\ \mu}$. Indeed, at the background level, the right-hand side of Eq.~(\ref{eq:conservation_einstein_matter}) vanishes for nonrelativistic particles ($T^0_{\ 0} = T$) \footnote{{The fact that the hatted energy momentum tensor satisfies the standard conservation law in background cosmology makes it more straightforward to calculate its background density evolution.}}. At the perturbed level, however, $\hat{T}_{\mu\nu}$ is not conserved. The interpretation of this is that dark matter and baryon particles feel a fifth force and a frictional force induced by the scalar coupling (see Secs.~\ref{subsubsec:_matter_quations_cdm} and \ref{subsubsec:_matter_quations_cdm} below).


In what follows, we will neglect the overhat of $\hat{T}^\mu_{\ \nu}$ and on its components when referring to baryons and cold dark matter, to lighten our notation.  However, bear in mind that when $\hat{T}^\mu_{\ \nu}$ (or $T^\mu_{\ \nu}$ hereafter) enters the {\it Einstein equation} and the {\it scalar field equation} Eq.~(\ref{eq:scalar_eom}), it should always be multiplied by an extra factor of $A(\varphi)$ compared with the $T^\mu_{\ \nu}$ for other matter species.


\subsection{The scalar field}

\label{subsect:scalar_eqns}

The scalar field Lagrangian in Eq.~(\ref{eq:action}) is already written in the Einstein frame, and thus there is no need to change frames. In the case of the K-mouflage field, the Lagrangian is purely kinetic:
\begin{eqnarray}\label{eq:scalar_lagrangian}
\mathcal{L}(\varphi) &=& -M^4K\left(\sigma\right)
\end{eqnarray}
in which $M$ is the characteristic mass scale of the model, and we have defined the dimensionless variable
\begin{eqnarray}
\sigma &\equiv& \frac{X}{M^4}\ \equiv\ \frac{1}{2M^4}\nabla^\rho\varphi\nabla_\rho\varphi,
\end{eqnarray}
to lighten the notation.

By varying the action in Eq.~(\ref{eq:action}) with respect to the scalar field $\varphi$, one obtains the K-mouflage equation of motion
\begin{eqnarray}\label{eq:scalar_eom}
\nabla_\lambda\left[K_\sigma(\sigma)\nabla^\lambda\varphi\right] &=& - \frac{{\rm d}\ln A(\varphi)}{{\rm d}\varphi}T^{(\nu)} - \sum_{i=c,b}\frac{{\rm d}A(\varphi)}{{\rm d}\varphi} T^{(i)}, \ \ \ \ \ \
\end{eqnarray}
in which the subscript $\sigma$ denotes partial derivative w.r.t. $\sigma$ and $T^{(\nu,c,b)}$ are, respectively, the traces of the energy momentum tensors for massive neutrinos, cold dark matter and baryons. As photons and massless neutrinos are traceless, they do not contribute to this equation. Note that, as we mentioned at the end of the previous subsection: $T^{(c,b)}$ should be understood as the hatted quantities, and so is multiplied by an extra factor of $A(\varphi)$ compared with $T^{(\nu)}$, which is why it has a coefficient of ${\rm d}A/{\rm d}\varphi$ instead of $\rm d \ln A/ \rm d \varphi$.

\section{Linear perturbation equations}

\label{sect:pert_equations}

In this section, applying the method of $3+1$ \cite{cl1999} space-time decomposition, we derive and summarise the fully covariant and gauge invariant (CGI) linearly perturbed equations in the K-mouflage model. We shall first present the general formalism of the $3+1$ decomposition, and then focus on the perturbations of the individual matter species. We pay particular attention to the perturbations of the massive neutrinos, which we believe have not been thoroughly explored in the past. We hope that the treatment outlined in this section can serve as a useful reference for future works.

\subsection{The $3+1$ decomposition}

\label{subsect:decomposition}

The idea of $3+1$ decomposition is to make spacetime splits of physical quantities with respect to an observer's 4-velocity, $u^\mu$. One can define a projection tensor $h_{\mu\nu}$ as
\begin{eqnarray}
h_{\mu\nu} &\equiv& g_{\mu\nu} - u_\mu u_\nu,
\end{eqnarray}
which can be used to obtain covariant tensors which reside on 3-dimensional hyperspaces perpendicular to $u^\mu$. For example, the covariant spatial derivative $\hat{\nabla}^\alpha$ of a tensor field, $T^{\beta...\gamma}_{\sigma...\lambda}$, is defined by the following relation
\begin{eqnarray}
\hat{\nabla}^\alpha T^{\beta...\gamma}_{\sigma...\lambda} &\equiv& h^{\alpha}_{\mu}h^{\beta}_{\nu}...h^{\gamma}_{\kappa}h^{\rho}_{\sigma}...h^{\eta}_{\lambda}\nabla^\mu T^{\nu...\kappa}_{\rho...\eta}.
\end{eqnarray}

Using this, the general energy-momentum tensor of matter and covariant derivative of the 4-velocity can be split, respectively, as
\begin{eqnarray}
\label{Tuv} T_{\mu\nu} &=& \pi_{\mu\nu} + 2q_{(\mu}u_{\nu)} + \rho u_\mu u_\nu - Ph_{\mu\nu},\\
\nabla_\mu u_\nu &=& \sigma_{\mu\nu} + \varpi_{\mu\nu} + \frac{1}{3}\theta h_{\mu\nu} + u_\mu w_\nu,
\end{eqnarray}
in which $\pi_{\mu\nu}$ is the projected symmetric and trace-free (PSTF) anisotropic stress, $q_\mu$ is the heat flux vector, $P$ is the isotropic pressure, $\sigma_{\mu\nu}$ the PSTF shear tensor, $\varpi_{\mu\nu} \equiv \hat{\nabla}_{[\mu}u_{\nu]}$ the antisymmetric vorticity tensor, $\theta \equiv \nabla^\alpha u_\alpha $ the expansion scalar\footnote{{$\theta$ is the full expansion scalar, and contains a background part and perturbations around it. To lighten notations, in this paper we use the same symbol $\theta$ for both the full quantity and its background part. Since we are only interested in linear perturbations, the exact meaning should be clear in a given context -- for example, when it is multiplied by a perturbation quantity then $\theta$ means only its background part.}} and $w_\mu \equiv \dot{u}_\mu$; the overdot denotes a time derivative defined as $\dot{\phi} = u^\alpha\nabla_\alpha\phi$, square brackets denote antisymmetrisation and parentheses symmetrisation {of indices (below we will also use angle brackets to denote symmetrisation and removal of the trace part)}. The normalisation is such that $u^\alpha u_\alpha = 1$, which is consistent with our metric sign convention $(+,-,-,-)$. The quantities $\pi_{\mu\nu}$, $q_\mu$, $\rho$ and $P$ are usually called dynamical quantities and $\sigma_{\mu\nu}$, $\varpi_{\mu\nu}$, $\theta$ and $w_\mu$ are called kinematical quantities. The dynamical quantities can be derived from the energy momentum tensor $T_{\mu\nu}$, Eq.~(\ref{Tuv}), as
\begin{eqnarray}\label{Tuv projection}
\rho &=& T_{\mu\nu}u^\mu u^\nu, \nonumber \\
P &=& -\frac{1}{3}h^{\mu\nu}T_{\mu\nu}, \nonumber \\
q_\mu &=& h_\mu^\nu u^\rho T_{\nu\rho}, \nonumber \\
\pi_{\mu\nu} &=& h_\mu^\rho h_\nu^\tau T_{\rho\tau} + Ph_{\mu\nu}.
\end{eqnarray}

\subsection{Einstein equations}

\label{subsect:einstein_equations}

The Einstein field equations can also be split in the $3+1$ framework \cite{cl1999},  to obtain a set of five propagation equations (those which govern the time evolution of perturbation variables) and five constraint equations (those which specify the relations between different perturbation variables). The structure of Einstein equations,
\begin{eqnarray}
G_{\mu\nu} &=& \kappa T_{\mu\nu},
\end{eqnarray}
{with $\kappa\equiv8\pi G$,} holds for all models, such as the K-mouflage model, as long as the extra terms are properly absorbed in $T_{\mu\nu}$. Decomposing the Riemann tensor, and after linearisation, the five constraint equations are given by
\begin{eqnarray}\label{c1}
0 &=& \hat{\nabla^\alpha}\left( \epsilon^{\mu\nu}_{\ \ \ \alpha\beta}u^\beta\varpi_{\mu\nu}\right), \\
\label{c2}
\kappa q_\mu &=& -\frac{2\hat{\nabla}_\mu\theta}{3} + \hat{\nabla}^\nu\sigma_{\mu\nu} + \hat{\nabla}^\nu\varpi_{\mu\nu},\\
\label{c3}
\mathcal{B}_{\mu\nu} &=& \left[ \hat{\nabla}^\alpha\sigma_{\beta(\mu} + \hat{\nabla}^\alpha\varpi_{\beta(\mu}\right]\epsilon_{\nu)\gamma\alpha}^{\ \ \ \ \beta} u^\gamma, \\
\label{c4}
\hat{\nabla}^\nu\mathcal{E}_{\mu\nu} &=& \frac{1}{2}\kappa\left[ \hat{\nabla}\pi_{\mu\nu} + \frac{2}{3}\theta q_\mu +  \frac{2}{3}\hat{\nabla}_\mu\rho\right], \\
\label{c5}
\hat{\nabla}^\nu\mathcal{B}_{\mu\nu} &=& \frac{1}{2}\kappa\left[ \hat{\nabla}_\alpha q_\beta + (\rho + P)\varpi_{\alpha\beta}\right]\epsilon_{\mu\nu}^{\ \ \alpha\beta}u^\nu,
\end{eqnarray}
whereas the five propagation equations are given by
\begin{eqnarray}\label{p1}
0 &=& \dot{\theta} + \frac{1}{3}\theta^2 - \hat{\nabla} \cdot w + \frac{\kappa}{2}(\rho + 3P), \\
\label{p2}
0 &=& \dot{\sigma}_{\mu\nu} + \frac{2}{3}\theta\sigma_{\mu\nu} - \hat{\nabla}_{\langle\mu}w_{\nu\rangle} + \mathcal{E}_{\mu\nu} + \frac{k}{2}\pi_{\mu\nu}, \\
\label{p3}
0 &=&\dot{\varpi}_{\mu\nu} + \frac{2}{3}\theta\varpi_{\mu\nu} - \hat{\nabla}_{[\mu}w_{\nu]}, \\
\label{p4}
0 &=& \frac{\kappa}{2}\left[ \dot{\pi}_{\mu\nu} + \frac{1}{3}\theta\pi_{\mu\nu}\right] - \frac{\kappa}{2}\left[ (\rho + P)\sigma_{\mu\nu} + \hat{\nabla}_{\langle\mu}q_{\nu\rangle}\right] \nonumber \\
&& - \left[ \dot{\mathcal{E}}_{\mu\nu} + \theta\mathcal{E}_{\mu\nu} - \hat{\nabla}^\alpha\mathcal{B}_{\beta(\mu}\epsilon_{\nu)\gamma\alpha}^{\ \ \ \ \ \beta}u^\gamma\right], \\
\label{p5}
0 &=&\dot{\mathcal{B}}_{\mu\nu} + \theta\mathcal{B}_{\mu\nu} + \hat{\nabla}^\alpha\mathcal{E}_{\beta(\mu}\epsilon_{\nu)\gamma\alpha}^{\ \ \ \ \ \beta}u^\gamma \\
&&+ \frac{\kappa}{2}\hat{\nabla}^\alpha\pi_{\beta(\mu}\epsilon_{\nu)\gamma\alpha}^{\ \ \ \ \ \beta}u^\gamma.
\end{eqnarray}
In these equations, $\epsilon_{\mu\nu\alpha\beta}$ is the 4-dimensional covariant permutation tensor, $\hat{\nabla} \cdot w \equiv \hat{\nabla}^\alpha w_\alpha$ (for any vector $w_\alpha$), and $\mathcal{E}_{\mu\nu}$ and $\mathcal{B}_{\mu\nu}$ are, respectively, the electric and magnetic parts of the Weyl tensor, $\mathcal{W}_{\mu\nu\alpha\beta}$, defined by $\mathcal{E}_{\mu\nu} \equiv u^\alpha u^\beta\mathcal{W}_{\mu\alpha\nu\beta}$ and $\mathcal{B}_{\mu\nu} \equiv -\frac{1}{2}u^\alpha u^\beta \epsilon_{\mu\alpha}^{\ \ \gamma\delta}\mathcal{W}_{\gamma\delta\nu\beta}$.

In addition to the above equations, it is often useful to express the projected Ricci scalar, $\hat{R}$, onto the hypersurfaces orthogonal to $u^\mu$, as
\begin{eqnarray}\label{spatial_curvature}
\hat{R} &=& 2\kappa\rho - \frac{2}{3}\theta^2.
\end{eqnarray}
The covariant spatial derivative of the projected Ricci scalar, $\eta_\mu \equiv a\hat{\nabla}_\mu\hat{R}/2$, can be derived from the above equation, as
\begin{eqnarray}\label{derivative_spatial_curvature}
\eta_\mu &=& \kappa a \hat{\nabla}_\mu\rho - \frac{2a}{3}\theta\hat{\nabla}_\mu\theta,
\end{eqnarray}
and its time evolution is governed by the following propagation equation
\begin{eqnarray}\label{propagation_spatial_curvature}
\dot{\eta}_\mu + \frac{2\theta}{3}\eta_\mu &=& -\frac{2a\theta}{3}\hat{\nabla}_\mu\hat{\nabla}\cdot w - a\kappa\hat{\nabla}_\mu\hat{\nabla}\cdot q.
\end{eqnarray}

The total energy-momentum tensor satisfies the conservation (continuity and Euler) equations,
\begin{eqnarray}\label{conservation1}
\dot{\rho} + (\rho + P)\theta + \hat{\nabla}\cdot q &=& 0, \\
\label{conservation2}
\dot{q}_\mu + \frac{4}{3}\theta q_{\mu} + (\rho+P)w_\mu - \hat{\nabla}_\mu P + \hat{\nabla}^\nu\pi_{\mu\nu} &=& 0.
\end{eqnarray}

In this paper, we focus on spatially-flat cosmologies, for which the spatial curvature vanishes at the background level, $\hat{R} = 0$. Therefore, from Eq.~(\ref{spatial_curvature}), we obtain the first Friedmann equation
\begin{eqnarray}\label{background1}
\frac{\theta^2}{3} &=& \kappa {\rho}.
\end{eqnarray}
Recall that at the background level only the zeroth-order terms contribute to the equations. The second Friedmann equation and the energy-conservation equation can be obtained by taking the zeroth-order parts of Eqs.~(\ref{p1}, \ref{conservation1}), as
\begin{eqnarray}\label{background2}
\dot{\theta} + \frac{1}{3}\theta^2 + \frac{\kappa}{2}({\rho} + 3{P}) &=& 0, \\
\label{background3}
\dot{{\rho}} + ({\rho} + {P})\theta &=& 0,
\end{eqnarray}


For the purpose of facilitating the numerical calculations, it is customary to work in $k$-space (Fourier space), in which spatial partial derivatives can be replaced with products of $k$ (here $k$ denotes the wave number of a given perturbation). This also simplifies the equations. To move the equations to Fourier space we use the following harmonic expansions for the perturbed quantities that enter the above equations:
\begin{widetext}
\begin{eqnarray}\label{eq:fouriermodes}
\mathcal{E}_{\mu\nu}\ =\ - \sum_k \frac{k^2}{a^2}\phi Q_{\mu\nu}^k, \quad \hat{\nabla}_\mu\theta\ =\ \sum_k\frac{k^2}{a^2}\mathcal{Z}Q_\mu^k,\quad \eta_\mu\ =\ \sum_k \frac{k^3}{a^2}\eta Q_\mu^k,\quad
w_\mu\ =\ \sum_k\frac{k}{a}wQ_\mu^k,\quad \sigma_{\mu\nu}\ =\ \sum_k \frac{k}{a}\sigma Q_{\mu\nu}^k,  \nonumber \\
\hat{\nabla}_\mu\rho\ =\ \sum_k \frac{k}{a}\mathcal{X} Q_\mu^k,\quad \hat{\nabla}_\mu P\ =\ \sum_k \frac{k}{a}\mathcal{X}^pQ_\mu^k,\quad {q}_{\mu}\ =\ \sum_k q Q_{\mu}^k,\quad {\pi}_{\mu\nu}\ =\ \sum_k \Pi Q_{\mu\nu}^k,\quad  \hat{\nabla}_\mu\varphi\ =\ \sum_k \frac{k}{a}\xi Q_\mu^k,\quad
\end{eqnarray}
\end{widetext}
where $Q^k$ is the eigenfunction of the comoving spatial Laplacian $a^2\hat{\Box}$ operator, which satisfies
\begin{eqnarray}
\hat{\Box}Q^k = \frac{k^2}{a^2}Q^k,
\end{eqnarray}
and $Q_\mu^k$ and $Q_{\mu\nu}^k$ are respectively defined by $Q_\mu^k \equiv \frac{a}{k}\hat{\nabla}_\mu Q^k$ and $Q_{\mu\nu}^k \equiv \frac{a}{k}\hat{\nabla}_{\langle\mu}Q_{\nu\rangle}$. We represent the scalar field perturbation mode in Fourier space as $\xi$, not $\delta\varphi$, to highlight the fact that this has been obtained in a covariant way {(typically, $\delta\varphi$ is used in the literature to denote gauge-{\it non}invariant scalar field perturbations).}

In terms of these harmonic expansion variables, Eqs.~(\ref{c2}, \ref{c4}, \ref{p2}, \ref{p4}, \ref{derivative_spatial_curvature}, \ref{propagation_spatial_curvature}) can be rewritten as
\begin{eqnarray}
\label{einsteink1}\frac{2}{3}k^2(\sigma - \mathcal{Z}) &=& \kappa qa^2, \\
\label{einsteink2}k^3\phi &=& -\frac{1}{2}\kappa a^2 \left[ k(\Pi + \chi) + 3\mathcal{H}q \right], \\
\label{einsteink3}k(\sigma' + \mathcal{H}\sigma) &=& k^2(\phi + w) - \frac{1}{2}\kappa a^2\Pi, \\
\label{einsteink4}k^2(\phi' + \mathcal{H}\phi) &=& \frac{1}{2}\kappa a^2 \left[ k(\rho + P)\sigma + kq - \Pi' - \mathcal{H}\Pi \right], \\
\label{einsteink5}k^2\eta &=& \kappa \chi a^2 - 2k\mathcal{H}\mathcal{Z}, \\
\label{einsteink6}k\eta' &=& -\kappa q a^2 - 2k\mathcal{H}w,
\end{eqnarray}
in which $\mathcal{H} \equiv a'/a$ and a prime represents the derivative with respect to the conformal time $\tau$ ($ad\tau = dt$, with $t$ the physical time).

Similarly, the conservation equations, Eqs.~(\ref{conservation1}, \ref{conservation2}), can be written in $k$-space as,
\begin{eqnarray}
\label{conservationk1}
\chi' + (k\mathcal{Z} - 3\mathcal{H}w)(\rho + P) + 3\mathcal{H}(\mathcal{X} + \mathcal{X}^p) + kq &=& 0,\ \ \ \ \ \\
\label{conservationk2}
q' + 4\mathcal{H}q + (\rho + P)kw - k\mathcal{X}^p + \frac{2}{3}k\Pi &=& 0.
\end{eqnarray}

{In the numerical implementation of the {\sc Camb} code we use for this study, Eqs.~(\ref{conservationk1}, \ref{conservationk2}) are solved numerically for individual matter species to compute $\chi$ and $q$ for those species at any given time (for photons and neutrinos we solve the Boltzmann hierarchies instead of solving these equations directly; see discussions below). Then the total $\chi$ and $q$ are used in Eqs.~(\ref{einsteink1} - \ref{einsteink6}) to solve the curvature variables $\mathcal{Z}$, $\phi$, $\sigma$ and $\eta$ ($w$ will be set to 0 as a choice of frame). The anisotropic stress $\Pi$ receives contributions from photons and neutrinos (both massless and massive), and can be computed using the integration of Eq.~(\ref{eq:neutrino_emt_3}).}

\subsection{Perturbation equations of the individual species}

\label{subsect:matter_equations}

In this subsection we present the linear perturbation equations for the evolution of each of the individual matter species that we consider: cold dark matter, baryons, photons, massless and massive neutrinos and the K-mouflage field.

\subsubsection{Cold dark matter}\label{subsubsec:_matter_quations_cdm}

{The cold dark matter fluid is collisionless, and as a result,
generates no pressure or anisotropic stress.
Hence, one obtains the energy-momentum tensor for cold dark matter as}

\begin{eqnarray}
 T^{(c)}_{\mu\nu} &=& \rho^{(c)}u_{\mu}u_{\nu} + 2u_{(\mu}q^{(c)}_{\nu)},
\end{eqnarray}
in which  we have omitted the "hat" on the different components of the tensor defined in Eq.~(\ref{eq:hat_T_munu}). By applying Eq.~(\ref{eq:conservation_einstein_matter}) and keeping the terms that are parallel to the 4-velocity of the observer, we obtain the continuity equation
\begin{eqnarray}\label{eq:perturbed_cdm_continuity}
\dot{\rho}^{(c)} + \theta\rho^{(c)} + \hat{\nabla}\cdot q^{(c)} &=& 0,
\end{eqnarray}
whose background part gives the usual energy conservation equation
\begin{eqnarray}\label{eq:background_cdm_continuity}
\dot{\bar{\rho}}^{(c)} + 3H\bar{\rho}^{(c)} &=& 0.
\end{eqnarray}
The terms perpendicular to the observer's 4-velocity give the Euler equation up to linear order:
\begin{eqnarray}\label{eq:perturbed_cdm_euler}
\dot{q}^{(c)}_\mu + \frac{4}{3}\theta q^{(c)}_{\mu} +{\rho}^{(c)}w_\mu + \frac{{\rm d}\ln A(\varphi)}{{\rm d}\varphi}\dot{\varphi}q^{(c)}_\mu && \nonumber\\
- \frac{{\rm d}\ln A(\varphi)}{{\rm d}\varphi}\rho^{(c)}\hat{\nabla}_\mu\varphi && \ =\ 0,
\end{eqnarray}
in which we have dropped the overbars for $\theta, \rho^{(c)}$ and $\dot{\varphi}$, since the context determines that these are background quantities.

The fact that cold dark matter satisfies the standard continuity equation, at both the background (cf.~Eq.~(\ref{eq:background_cdm_continuity})) and the linear perturbation (cf.~Eq.~(\ref{eq:perturbed_cdm_continuity})) level, is a consequence of the redefinition of the energy-momentum tensor in Eq.~(\ref{eq:hat_T_munu}). However, even after this redefinition, cold dark matter particles do not satisfy the standard Euler equation. Instead, they experience an additional "fifth" force, as determined by the last term of Eq.~(\ref{eq:perturbed_cdm_euler}). In addition to the fifth force, the scalar coupling to the cold dark matter particles induces also an additional friction term, as represented by the second last term. This implies changes in the dark matter perturbation evolution.

In $k$-space, the continuity and Euler equations can be written as
\begin{eqnarray}
\label{eq:k_continuity_c} \Delta'_{(c)} + k\mathcal{Z} - 3\frac{a'}{a}w + kv_{(c)} &=& 0,\\
\label{eq:k_euler_c} v'_{(c)} + \frac{a'}{a}v_{(c)} + kw + \frac{{\rm d}\ln A(\varphi)}{{\rm d}\varphi}\left(\varphi'v_{(c)}-k\xi\right) &=& 0,
\end{eqnarray}
in which we have defined $q^{(c)}\equiv\bar{\rho}^{(c)}v_{(c)}$ and $\Delta_{(c)}$ is the density contrast for cold dark matter. {Eqs.~(\ref{eq:k_continuity_c}, \ref{eq:k_euler_c}) are directly used in our modified {\sc Camb} code.}

\subsubsection{Baryons}\label{subsubsec:matter_equations_baryons}

Baryons are similar to cold dark matter, with the difference that they produce a non-negligible pressure at the linear perturbation level. The pressure, however, vanishes on the cosmological background. The energy momentum tensor is then given by
\begin{eqnarray}
 T^{(b)}_{\mu\nu} &=& \rho^{(b)}u_{\mu}u_{\nu} - P^{(b)}h_{\mu\nu} + 2u_{(\mu}q^{(b)}_{\nu)},
\end{eqnarray}
and following the same steps as for cold dark matter we obtain the continuity equation
\begin{eqnarray}
\dot{\rho}^{(b)} + \theta\left(\rho^{(b)}+P^{(b)}\right) + \hat{\nabla}\cdot q^{(b)} &=&  -3 P^{(b)} \frac{ {\rm d} \ln A(\varphi)}{{\rm d} \varphi} \dot \varphi, ~~~ ~~~
\end{eqnarray}
and the modified Euler equation
\begin{eqnarray}
\dot{q}^{(b)}_\mu + \frac{4}{3}\theta q^{(b)}_{\mu} + {\rho}^{(b)}w_\mu - \hat{\nabla}_{\mu}P^{(b)} && \nonumber\\
+ \frac{{\rm d}\ln A(\varphi)}{{\rm d}\varphi}\dot{\varphi}q^{(b)}_\mu - \frac{{\rm d}\ln A(\varphi)}{{\rm d}\varphi}\rho^{(b)}\hat{\nabla}_\mu\varphi &&\ =\ 0,
\end{eqnarray}
up to linear order in real space.  We have neglected terms such as $P^{(b)}w_\mu$ and $P^{(b)}\hat{\nabla}_\mu\varphi$, which are higher than first order in perturbations because $P^{(b)}$ is a perturbed quantity (i.e. $\bar{P}^{(b)} = 0$).   On the cosmological background, taking the zeroth-order terms, we obtain the standard energy conservation equation
\begin{eqnarray}\label{eq:background_continuity_b}
\dot{\bar{\rho}}^{(b)} + 3H\bar{\rho}^{(b)} &=&0 .
\end{eqnarray}

In $k$-space, the continuity and Euler equations for baryons become, respectively,
\begin{eqnarray}
\label{eq:k_continuity_b} \Delta'_{(b)} + k\mathcal{Z} - 3  \frac{a'}{a} w + kv_{(b)}  & & \nonumber \\
+ 3 \left( \frac{a'}{a} + \frac{ {\rm d} \ln A(\varphi)}{{\rm d} \varphi } \varphi' \right)   c_s^2\Delta_{(b)} &=& 0,\\
\label{eq:k_euler_b} v'_{(b)} + \frac{a'}{a}v_{(b)} + kw - kc_s^2\Delta_{(b)} &&\nonumber\\
 + \frac{{\rm d}\ln A(\varphi)}{{\rm d}\varphi}\left(\varphi'v_{(b)}-k\xi\right) &=& 0,
\end{eqnarray}
in which we have defined the baryon sound speed squared $c_s^2$ as $c_s^2\equiv\mathcal{X}^{p,(b)}/\mathcal{X}^{(b)}$, and $v_{(b)}$, $\Delta_{(b)}$ are, respectively, the peculiar velocity and density contrast of baryons.

Note that the above Euler equation is derived assuming no interaction between baryons and photons. To account for this, we simply add the term that describes the Thomson scattering to the equation to get
\begin{eqnarray}\label{eq:k_euler_b2}
 0 &  = &  v'_{(b)} + \frac{a'}{a}v_{(b)} + kw - kc_s^2\Delta_{(b)}   \nonumber\\
& + &  \frac{{\rm d}\ln A(\varphi)}{{\rm d}\varphi}\left(\varphi'v_{(b)}-k\xi\right) \nonumber\\
& + &    a n_e\sigma_{\rm T}\frac{\bar{\rho}^{(\gamma)}}{\bar{\rho}^{(b)}}\left(\frac{4}{3}v_{(b)}-v_{(\gamma)}\right),
\end{eqnarray}
in which $n_e$ is the free electron number density, $\sigma_{\rm T}$ is the cross section for Thomson scattering, $v_{(\gamma)}$ is the peculiar velocity for the photon fluid, and $\bar{\rho}^{(\gamma)}$ is the background energy density of photons. {Eqs.~(\ref{eq:k_continuity_b}, \ref{eq:k_euler_b}) are used our modified {\sc Camb} code; when the Thomson scattering is not negligible Eq.~(\ref{eq:k_euler_b2}) is used instead of Eq.~(\ref{eq:k_euler_b}).}

\subsubsection{Photons}

As we have found in Sec.~\ref{sect:equations}, the photon Lagrangian density is conformally invariant, and the photon energy momentum tensor is conserved in the Einstein frame (the frame where we perform our calculations). Consequently, the evolution of the photon fluid is the same as it would be in the case of the absence of the scalar coupling. Nevertheless, for completeness, we simply note that the first two moments of the angular expansion of the photon distribution function lead to the following continuity and Euler equations in real space \cite{cl1999}
\begin{eqnarray}
\dot{\rho}^{(\gamma)} + \frac{4}{3}\theta\rho^{(\gamma)} + \hat{\nabla}\cdot q^{(\gamma)} &=& 0,\ \ \ \\
\dot{q}^{(\gamma)}_{\mu} + \frac{4}{3}\theta q^{(\gamma)}_{\mu} + \frac{4}{3}\rho^{(\gamma)}w_{\mu} - \frac{1}{3}\hat{\nabla}_{\mu}\rho^{(\gamma)} + \hat{\nabla}^{\nu}\pi^{(\gamma)}_{\mu\nu}\nonumber\\
-n_e\sigma_{\rm T}\left(\frac{4\rho^{(\gamma)}}{3\rho^{(b)}}q^{(b)}_\mu - q^{(\gamma)}_\mu\right) &=& 0,
\end{eqnarray}
in which we have added the Thomson scattering term. As expected, at the background level, we recover the standard radiation conservation equation
\begin{eqnarray}\label{eq:background_continuity_p}
\dot{\bar{\rho}}^{(\gamma)}  + 4H\bar{\rho}^{(\gamma)} &=& 0.
\end{eqnarray}

In $k$ space the perturbed continuity and Euler equations become
\begin{eqnarray}
\label{eq:k_continuity_p} \Delta'_{(\gamma)} + \frac{4}{3}k\mathcal{Z} - 4\frac{a'}{a}w + kv_{(\gamma)} &=& 0,\\
\label{eq:k_euler_p} v'_{(\gamma)} + \frac{4}{3}kw - \frac{1}{3}k\Delta_{(\gamma)} + \frac{2}{3}k\pi_{(\gamma)}\nonumber\\
 + an_e\sigma_{\rm T}\left(v_{(\gamma)}-\frac{4}{3}v_{(b)}\right) &=& 0,
\end{eqnarray}
in which $v_{(\gamma)}\equiv q^{(\gamma)}/\bar{\rho}^{(\gamma)}$,  $\pi_{(\gamma)}\equiv\Pi^{(\gamma)}/\bar{\rho}^{(\gamma)}$ ($\pi$ here should not be confused with the real space quantity $\pi_{\mu\nu}$ in Eq.~(\ref{Tuv})) and $\Delta_{(\gamma)}$ is the density contrast of photons. {Eqs.~(\ref{eq:k_continuity_p}, \ref{eq:k_euler_p}) are used directly in our modified {\sc Camb} code.} 

For brevity, the higher order moments of the angular expansion of the photon distribution function are not shown here.

\subsubsection{Massless neutrinos}\label{subsubsec:massless_nu}

Massless neutrinos are very similar to photons, except that they do not interact with the baryons via Thomson scattering. The real space continuity and Euler equations are therefore given by
\begin{eqnarray}
\dot{\rho}^{(r)} + \frac{4}{3}\theta\rho^{(r)} + \hat{\nabla}\cdot q^{(r)} &=& 0,\ \ \ \\
\dot{q}^{(r)}_{\mu} + \frac{4}{3}\theta q^{(r)}_{\mu} + \frac{4}{3}\rho^{(r)}w_{\mu} - \frac{1}{3}\hat{\nabla}_{\mu}\rho^{(r)} + \hat{\nabla}^{\nu}\pi^{(r)}_{\mu\nu}  &=& 0.
\end{eqnarray}
We use $(r)$ to denote massless neutrinos quantities, to distinguish them from the massive neutrinos ones, which we denote by ($\nu$). At the cosmological background level, the energy density satisfies the usual radiation conservation equation
\begin{eqnarray}\label{eq:background_continuity_r}
\dot{\bar{\rho}}^{(r)}  + 4H\bar{\rho}^{(r)} &=& 0.
\end{eqnarray}

In $k$ space, the perturbed continuity and Euler equations become
\begin{eqnarray}
\label{eq:k_continuity_r} \Delta'_{(r)} + \frac{4}{3}k\mathcal{Z} - 4\frac{a'}{a}w + kv_{(r)} &=& 0,\\
\label{eq:k_euler_r} v'_{(r)} + \frac{4}{3}kw - \frac{1}{3}k\Delta_{(r)} + \frac{2}{3}k\pi_{(r)} &=& 0,
\end{eqnarray}
in which $v_{(r)}\equiv q^{(r)}/\bar{\rho}^{(r)}$, $\pi_{(r)}\equiv\Pi^{(r)}/\bar{\rho}^{(r)}$, and $\Delta_{(r)}$ is the density perturbation of massless neutrinos. {Eqs.~(\ref{eq:k_continuity_r}, \ref{eq:k_euler_r}) are used directly our modified {\sc Camb} code.} 

For brevity, we shall not show the full hierarchy of equations satisfied by the higher-order angular moments of the massless neutrinos distribution function.

\subsubsection{Massive neutrinos}

The case for massive neutrinos is slightly subtler. For typical masses within the allowed observational bounds \cite{gc2009, gd2013}, the equation-of-state parameter of massive neutrinos evolves from $w_{(\nu)} = 1/3$ at earlier times (when they are highly relativistic) to $w_{(\nu)} \approx 0$ at later times (once they become nonrelativistic). For this reason, we can not simply redefine their energy momentum tensor such that $\bar{\rho}^{(\nu)}$ evolves as a power-law function of the scale factor, $a$ (as we have done for cold dark matter and baryons). However, this fact does not make the computation of $\bar{\rho}^{(\nu)}$ in the K-mouflage model any more complicated. Indeed, it is straightforward to check that, at the background level, Eqs.~(\ref{eq:neutrino_emt_1}, \ref{eq:neutrino_emt_3}) satisfy the conservation equations, Eq.~(\ref{eq:conservation_einstein}), provided one takes into account the fact that the mass, $m$, in these equations, is time varying. The evolution $\bar{\rho}^{(\nu)}$ is normally computed by working out the integrals in Eq.~(\ref{eq:neutrino_emt_1}) numerically. In our case, we do the same except that we must replace $m$ with $A(\varphi(a))\tilde{m}$, to account for the scalar coupling (where the bare neutrino mass $\tilde{m}$ is known). In this sense, the calculation of $\bar{\rho}^{(\nu)}$ in the K-mouflage model does not involve any more work compared to the standard uncoupled scenario, except for the operation of multiplying $\tilde{m}$ by $A(\varphi(a))$.

The linear perturbation evolution of massive neutrinos with no scalar coupling is well understood \cite{mb1995}. Although adding a nonminimal coupling is only a straightforward generalisation, for completeness, we shall nevertheless summarise the main steps of the derivation. Let us start with the geodesic equation of a point particle in the presence of the scalar coupling (in the Einstein frame),
\begin{eqnarray}\label{eq:particle_geodesic}
\frac{{\rm d}x^{\mu}}{{\rm d}s}\nabla_{\mu}\left[A(\varphi)\frac{{\rm d}x^{\nu}}{{\rm d}s}\right] &=& \frac{{\rm d}A(\varphi)}{{\rm d}\varphi}\nabla^{\nu}\varphi,
\end{eqnarray}
where ${\rm d}s$ is the proper length of the line element. In this equation, the $A(\varphi)$ factor on the left-hand side represents the varying particle mass (or an effective frictional force), while the $A(\varphi)$ factor on the right-hand side is responsible for the fifth force. In the case of non-relativistic particles, these terms correspond, respectively, to the extra friction and fifth force terms identified e.g. in Eq.~(\ref{eq:perturbed_cdm_euler}). For highly relativistic particles, on the other hand, the proper length vanishes, ${\rm d}s\rightarrow0$, and the geodesic equation reduces to:
\begin{eqnarray}
U^{\mu}\nabla_{\mu}U^{\nu} &=& 0,
\end{eqnarray}
as in the usual uncoupled case.

In the K-mouflage model, massive neutrinos can still be described by the collisionless Boltzmann equation in the eras we are interested in, but are subject to an external force due to the scalar coupling. Up to linear order in perturbed quantities, the Boltzmann equation is approximately given by
\begin{eqnarray}
\frac{\partial f}{\partial\tau} + \frac{{\rm d}x^i}{{\rm d}\tau}\frac{\partial f}{\partial x^i} + \frac{{\rm d}q}{{\rm d}\tau}\frac{\partial f}{\partial q} &=& 0,
\end{eqnarray}
where $q$ is the magnitude of neutrino momentum (not the heat flux), and remember that $f=f({\bf x}, q, {\bf n}, \tau)$. Using Eq.~(\ref{eq:particle_geodesic}) to replace ${\rm d}q/{\rm d}\tau$ in this equation, and moving to $k$ space, we reach the following evolution equation for $\Psi({\bf x}, q, {\bf n}, \tau)$:
\begin{eqnarray}
\Psi' + i\left(\hat{k}\cdot\hat{n}\right)\frac{q}{\epsilon}k\Psi\nonumber\\
+ \frac{{\rm d}\ln f_0(q)}{{\rm d}\ln q}\left[\left(\frac{1}{3}k\sigma-h'\right)-\left(\hat{k}\cdot\hat{n}\right)^2k\sigma\right]\nonumber\\
+ i\left(\hat{k}\cdot\hat{n}\right)\frac{{\rm d}\ln f_0(q)}{{\rm d}\ln q}k\left[\frac{{\rm d}\ln A(\varphi)}{{\rm d}\varphi}\frac{a^2m^2}{q\epsilon}\xi-\frac{\epsilon}{q}w\right] &=& 0,\ \ \ \ \ \ \ \ \
\end{eqnarray}
where $i=\sqrt{-1}$, $\hat{k}$ and $\hat{n}$ are respectively the unit vectors in the directions of ${\bf k}$ and ${\bf n}$, and
\begin{eqnarray}
h' &=& \frac{1}{3}k\mathcal{Z}-\frac{a'}{a}w
\end{eqnarray}
in a general frame.

To solve the above Boltzmann equation, one can expand the angular dependence of $\Psi$ in a series of Legendre polynomials $P_{\ell}\left(\hat{k}\cdot\hat{n}\right)$ as:
\begin{eqnarray}
\Psi({\bf k}, q, {\bf n}, \tau) &=& \sum_{\ell=0}^{\infty}(-i)^{\ell}(2\ell+1)\Psi_{\ell}({\bf k}, q, \tau)P_{\ell}\left(\hat{k}\cdot\hat{n}\right),\ \ \ \
\end{eqnarray}
so that each $\ell$-mode satisfies the following mode equation:
\begin{widetext}
\begin{eqnarray}\label{eq:boltzmann_series}
0 &=& \Psi'_{\ell} + \frac{k}{2\ell+1}\frac{q}{\epsilon}\left[(\ell+1)\Psi_{\ell+1}-\ell\Psi_{\ell-1}\right] + \frac{{\rm d}\ln f_0(q)}{{\rm d}\ln q}\left[\delta_{2\ell}\frac{2}{15}k\sigma - \delta_{1\ell}\frac{\epsilon}{3q}kw + \delta_{1\ell}\frac{{\rm d}\ln A(\varphi)}{{\rm d}\varphi}k\frac{a^2m^2}{3q\epsilon}\xi - \delta_{0\ell}h'\right],
\end{eqnarray}
\end{widetext}
in which $\delta_{0\ell}$, $\delta_{1\ell}$ and $\delta_{2\ell}$ are Kronecker $\delta$-functions. One can check this equation by verifying, with its help, that Eqs.~(\ref{eq:neutrino_emt_1} - \ref{eq:neutrino_emt_3}) satisfy the conservation equation, Eq.~(\ref{eq:conservation_einstein}), to linear order.

The $\delta_{0\ell}$ term in the Boltzmann equation makes sure that the local perturbations of the expansion rate \cite{lc2002} (recall that $h^{\prime}\propto\mathcal{Z}$, which is the $k$-space mode of $\hat{\nabla}\theta$) are properly taken into account in the calculation of the density contrast, which shall be expressed as an integral of $\Psi_0$ according to Eq.~(\ref{eq:emt_integrals}). The $\delta_{1\ell}$ terms contain the contributions from the scalar coupling and the 4-acceleration. Note how the equation reduces to that of massless neutrinos, when $m=0$ or $A(\varphi)=1$.

Eq.~(\ref{eq:boltzmann_series}) indicates that only the $\ell=1$ mode is affected by the scalar coupling. As $\ell=0, 1, 2$ contribute, respectively, to the energy density,  heat flux and anisotropic stress of neutrinos (see the definition and discussion of $I_{\ell}$ below), it might seem that the scalar coupling only changes the evolution of neutrino heat flux. We will show below that this is not the case, and that the density and pressure of massive neutrinos are also affected by the coupling. {Eq.~(\ref{eq:boltzmann_series}) is used in our modified {\sc Camb} code to solve the Boltzmann hierarchy of massive neutrinos accurately at early times, when the fluid approximation (discussed in the next paragraph) is not switched on.} 


At late times, neutrino momenta redshift away, and the so-called fluid approximation \cite{lc2002} (which involves considering only the modes $\ell \leq 2$) is often used to speed up the numerical calculations. The evolution equations under this approximation can be derived as follows. For each value of $\ell=0, 1, 2$ we (i) define
\begin{eqnarray}\label{eq:emt_integrals}
I_{\ell} &\equiv& \frac{4\pi}{\bar{\rho}^{(\nu)}a^4}\int{\rm d}qq^2\epsilon\left(\frac{q}{\epsilon}\right)^{\ell}f_0(q)\Psi_\ell,
\end{eqnarray}
(ii) multiply Eq.~(\ref{eq:boltzmann_series}) with $4\pi/\bar{\rho}^{(\nu)}a^4$, and {(iii)} integrate the resulting equation over ${\rm d}qq^2\epsilon(q/\epsilon)^{\ell}f_0(q)$. Doing so, we find
\begin{widetext}
\begin{eqnarray}
\label{eq:fluid_approx_1} I'_0 + \frac{a'}{a}\left(J_0-3\frac{\bar{P}^{(\nu)}}{\bar{\rho}^{(\nu)}}I_0\right) + kI_1 + 3\left(1+\frac{\bar{P}^{(\nu)}}{\bar{\rho}^{(\nu)}}\right)h' + \left(J_0-3\frac{\bar{P}^{(\nu)}}{\bar{\rho}^{(\nu)}}I_0\right)\frac{{\rm d}\ln A(\varphi)}{{\rm d}\varphi}\varphi' &=& 0,\\
\label{eq:fluid_approx_3} I'_1 + \frac{a'}{a}\left(1-3\frac{\bar{P}^{(\nu)}}{\bar{\rho}^{(\nu)}}\right)I_1 + \frac{2}{3}kI_2 - \frac{1}{3}kJ_0 + \left(1+\frac{\bar{P}^{(\nu)}}{\bar{\rho}^{(\nu)}}\right)kw \nonumber\\
+ \frac{{\rm d}\ln A(\varphi)}{{\rm d}\varphi}\left(1-3\frac{\bar{P}^{(\nu)}}{\bar{\rho}^{(\nu)}}\right)\varphi'I_1 - \frac{{\rm d}\ln A(\varphi)}{{\rm d}\varphi}\left(1-4\frac{\bar{P}^{(\nu)}}{\bar{\rho}^{(\nu)}}\right)k\xi  &=& 0,\\
\label{eq:fluid_approx_4} I'_2 + \frac{a'}{a}\left(2-3\frac{\bar{P}^{(\nu)}}{\bar{\rho}^{(\nu)}}\right)I_2 + \frac{3}{5}kI_3 - \frac{2}{5}kJ_1 - 2\frac{\bar{P}^{(\nu)}}{\bar{\rho}^{(\nu)}}k\sigma + \left(2-3\frac{\bar{P}^{(\nu)}}{\bar{\rho}^{(\nu)}}\right)\frac{{\rm d}\ln A(\varphi)}{{\rm d}\varphi}\varphi'I_2 &=& 0,
\end{eqnarray}
\end{widetext}
in which we have defined
\begin{eqnarray}
J_{\ell} &\equiv& \frac{4\pi}{\bar{\rho}^{(\nu)}a^4}\int{\rm d}qq^2\epsilon\left(\frac{q}{\epsilon}\right)^{2+\ell}f_0(q)\Psi_\ell,
\end{eqnarray}
whose evolution can be obtained by multiplying Eq.~(\ref{eq:boltzmann_series}) by $4\pi/\bar{\rho}^{(\nu)}a^4$ and integrating over ${\rm d}qq^2\epsilon(q/\epsilon)^{2+\ell}f_0(q)$:
\begin{widetext}
\begin{eqnarray}\label{eq:fluid_approx_4}
J'_0 + \frac{a'}{a}\left(2-3\frac{\bar{P}^{(\nu)}}{\bar{\rho}^{(\nu)}}\right)J_0 + kJ_1 + 15\frac{\bar{P}^{(\nu)}}{\bar{\rho}^{(\nu)}}h' + \left(2-3\frac{\bar{P}^{(\nu)}}{\bar{\rho}^{(\nu)}}\right)\frac{{\rm d}\ln A(\varphi)}{{\rm d}\varphi}\varphi'J_0  &=& 0.
\end{eqnarray}
\end{widetext}
Note that the fluid approximation is characterised by $\ell = 0, 1, 2$ for $I_{\ell}$, and $\ell = 0, 1$ for $J_{\ell}$. We note also that to ensure consistency we have used the following approximation:
\begin{eqnarray}
\int{\rm d}qf_0\frac{q^4}{\epsilon^3}a^2m^2 &\approx& \int{\rm d}qf_0\frac{q^4}{\epsilon}.\nonumber
\end{eqnarray}
The above equations do not reduce to the equations for the $\ell=0,1,2$ moments of massless neutrinos in the limit $m\rightarrow0$. This can be confirmed by observing that, if $\bar{P}^{(\nu)}\rightarrow\bar{\rho}^{(\nu)}/3$, then the terms that involve ${\rm d}\ln A/{\rm d}\varphi$ do not vanish, as they should for massless neutrinos (cf.~Sec.~\ref{subsubsec:massless_nu}). The reason for this is that the fluid approximation itself relies on the assumption that massive neutrinos are non-relativistic. Consequently, the fluid approximation breaks down in the case of massless neutrinos, for which $am=0<q$ instead of $am\gg q$. The case of standard massive neutrinos with no scalar coupling, on the other hand, can be recovered by setting $A(\varphi)=1$.

One may wonder whether or not $I_0, J_0, I_1, I_2$ can be identified with $\Delta_{(\nu)}, 3\mathcal{X}^{p,(\nu)}/\bar{\rho}^{(\nu)}$, $v_{(\nu)}$ and $\Pi_{(\nu)}/\bar{\rho}^{(\nu)}$ respectively, where $\Delta_{(\nu)}$ is the neutrino density contrast and $v_{(\nu)}$ the peculiar velocity. This is the case for standard uncoupled massive neutrinos. However, in the coupled case the answer is no, and one could check that the conservation equation, Eq.~(\ref{eq:conservation_einstein}), does not hold in this case if the above mapping is done. The reason lies in the spatial variation of the neutrino mass $m=A(\varphi)\tilde{m}$, whose contribution should be added to the components of the perturbed energy momentum. Doing so, the above quantities are then related as
\begin{eqnarray}
\Delta_{(\nu)} &=& I_0 + \frac{4\pi}{\bar{\rho}^{(\nu)}a^4}\int{\rm d}q\frac{q^2}{\epsilon}a^2m^2f_0\frac{{\rm d}\ln A(\varphi)}{{\rm d}\varphi}\xi\nonumber\\
&=& I_0 + \left(1-3\frac{\bar{P}^{(\nu)}}{\bar{\rho}^{(\nu)}}\right)\frac{{\rm d}\ln A(\varphi)}{{\rm d}\varphi}\xi,\label{eq:complete_delta_massnu}\\
\frac{\mathcal{X}^{p,(\nu)}}{\bar{\rho}^{(\nu)}} &=& \frac{1}{3}J_0 - \frac{\bar{P}^{(\nu)}}{\bar{\rho}^{(\nu)}}\frac{{\rm d}\ln A(\varphi)}{{\rm d}\varphi}\xi,\\
v_{(\nu)} &=& I_1,\\
\frac{\Pi_{(\nu)}}{\bar{\rho}^{(\nu)}} &=& I_2\ \equiv\  \pi_{(\nu)},\label{eq:complete_stress_massnu}
\end{eqnarray}
which are valid up to the linear order in perturbations and consistent with the fluid approximation. Note that the corrections to the heat flux and anisotropic stress due to the spatial variation of the neutrino mass are at most second order in perturbations and can therefore be neglected in our study.

Finally, it can be checked the completed variables, defined in Eqs.~(\ref{eq:complete_delta_massnu} - \ref{eq:complete_stress_massnu}), satisfy the conservation equations in the Einstein frame, Eq.~(\ref{eq:conservation_einstein}), up to linear order:
\begin{widetext}
\begin{eqnarray}
\Delta'_{(\nu)} + 3\left(\frac{a'}{a}+\frac{{\rm d}\ln A(\varphi)}{{\rm d}\varphi}\varphi'\right)\left(\frac{\mathcal{X}^{p,(\nu)}}{\bar{\rho}^{(\nu)}}-\frac{\bar{P}^{(\nu)}}{\bar{\rho}^{(\nu)}}\Delta_{(\nu)}\right) + \left(1+\frac{\bar{P}^{(\nu)}}{\bar{\rho}^{(\nu)}}\right)k\mathcal{Z} + kv_{(\nu)} - 3\frac{a'}{a}\left(1+\frac{\bar{P}^{(\nu)}}{\bar{\rho}^{(\nu)}}\right)w\nonumber\\
- \frac{{\rm d}\ln A(\varphi)}{{\rm d}\varphi}\left(1-3\frac{\bar{P}^{(\nu)}}{\bar{\rho}^{(\nu)}}\right)\xi' - \frac{{\rm d}^2\ln A(\varphi)}{{\rm d}\varphi^2}\left(1-3\frac{\bar{P}^{(\nu)}}{\bar{\rho}^{(\nu)}}\right)\varphi'\xi &=& 0,\\
v'_{(\nu)} + \frac{a'}{a}\left(1-3\frac{\bar{P}^{(\nu)}}{\bar{\rho}^{(\nu)}}\right)v_{(\nu)} - k\frac{\mathcal{X}^{p,(\nu)}}{\bar{\rho}^{(\nu)}} + \frac{2}{3}k\pi_{(\nu)} + \left(1+\frac{\bar{P}^{(\nu)}}{\bar{\rho}^{(\nu)}}\right)kw + \frac{{\rm d}\ln A(\varphi)}{{\rm d}\varphi}\left(1-3\frac{\bar{P}^{(\nu)}}{\bar{\rho}^{(\nu)}}\right)\left(\varphi'v_{(\nu)}-k\xi\right)  &=& 0.\ \ \
\end{eqnarray}
\end{widetext}
These equations reduce to those in the case of massless neutrinos when $\bar{P}^{(\nu)}\rightarrow\bar{\rho}^{(\nu)}/3$ and $\mathcal{X}^{p,(\nu)}\rightarrow\bar{\rho}^{(\nu)}\Delta_{(\nu)}/3$. Moreover, when $\bar{P}^{(\nu)}\rightarrow0$ and $\mathcal{X}^{p,(\nu)}\rightarrow0$ the Euler equation reduces to the case of non-relativistic matter, but note that this is not the case for the continuity equation due to our redefinition in Eq.~(\ref{eq:hat_T_munu}).

{Eqs.~(\ref{eq:fluid_approx_1} - \ref{eq:fluid_approx_4}) are used in our modified {\sc Camb} code when the fluid approximation is switched on at late times. Note that {\sc Camb} switches on this approximation at different times for different $k$-modes \cite{lc2002}, based on some rough estimate of when a given mode becomes non-relativistic. Also notice that other Boltzmann codes may treat the fluid approximation in slightly different ways, and special care needs to be taken if one is to use these equations in codes other than {\sc Camb}.}

\subsection{Scalar field equation}

\label{subsect:scalar_equation}

From the Lagrangian density of the K-mouflage field specified in Eq.~(\ref{eq:scalar_lagrangian}), one obtains the energy-momentum tensor for the scalar field as
\begin{eqnarray}
T^{(K)}_{\mu\nu} &=& K_\sigma\nabla_\mu\varphi\nabla_\nu\varphi - M^4Kg_{\mu\nu},
\end{eqnarray}
in which the superscript $^{(K)}$ stands for K-mouflage. Using Eqs.~(\ref{Tuv projection}), up to linear order, its components are given by
\begin{eqnarray}
\label{eq:sf_emt_com1}\rho^{(K)} &=& K_\sigma\dot{\varphi}^2-M^4K,\\
\label{eq:sf_emt_com2}P^{(K)} &=& M^4K,\\
\label{eq:sf_emt_com3}q^{(K)}_{\mu} &=& K_\sigma\dot{\varphi}\hat{\nabla}_\mu\varphi,\\
\label{eq:sf_emt_com4}\pi^{(K)}_{\mu\nu} &=& 0.
\end{eqnarray}
These reduce to the results of a quintessence scalar field with no potential when $K(\sigma)=\sigma$.

The background equation of motion of the scalar field can be read from Eq.~(\ref{eq:scalar_eom}) as
\begin{eqnarray}\label{eq:scalar_eom_background}
0 &=& \left(K_\sigma+2\sigma K_{\sigma\sigma}\right)\ddot{{\varphi}} + 3K_{\sigma}H\dot{\varphi}\\
&& +\frac{{\rm d}\ln A(\varphi)}{{\rm d}\varphi}\left(\bar{\rho}^{(\nu)}-3\bar{P}^{(\nu)}\right) + \frac{{\rm d}A(\varphi)}{{\rm d}\varphi}\left(\bar{\rho}^{(c)}+\bar{\rho}^{(b)}\right),\nonumber
\end{eqnarray}
where we have omitted the overbars on quantities such as $K$ and $\varphi$, to lighten the notation. {This equation is numerically solved in our modified {\sc Camb} code to obtain the background evolution.}

The perturbed equation of motion can be obtained by taking the covariant spatial derivative of Eq.~(\ref{eq:scalar_eom}), and we get
\begin{widetext}
\begin{eqnarray}\label{eq:scalar_eom_linear}
\left(K_{\sigma}+2\sigma K_{\sigma\sigma}\right)\hat{\nabla}_{\alpha}\ddot{\varphi} + \left(6\sigma K_{\sigma\sigma}+4\sigma^2K_{\sigma\sigma\sigma}\right)\frac{\ddot{\varphi}}{\dot{\varphi}}\hat{\nabla}_{\alpha}\dot{\varphi} + K_{\sigma}\theta\hat{\nabla}_{\alpha}\dot{\varphi} + K_{\sigma}\dot{\varphi}\hat{\nabla}_{\alpha}\theta + 2\sigma K_{\sigma\sigma}\theta\hat{\nabla}_{\alpha}\dot{\varphi} + K_{\sigma}\nabla_{\alpha}\hat{\nabla}^2\varphi\nonumber\\
+ \frac{{\rm d}\ln A(\varphi)}{{\rm d}\varphi}\left(\hat{\nabla}_{\alpha}{\rho}^{(\nu)}-3\hat{\nabla}_{\alpha}{P}^{(\nu)}\right) + \frac{{\rm d}A(\varphi)}{{\rm d}\varphi}\left(\hat{\nabla}_{\alpha}{\rho}^{(c)}+\hat{\nabla}_{\alpha}{\rho}^{(b)}\right)\nonumber\\
+ \frac{{\rm d}^2\ln A(\varphi)}{{\rm d}\varphi^2}\left(\bar{\rho}^{(\nu)}-3\bar{P}^{(\nu)}\right)\hat{\nabla}_{\alpha}\varphi + \frac{{\rm d}^2A(\varphi)}{{\rm d}\varphi^2}\left(\bar{\rho}^{(c)}+\bar{\rho}^{(b)}\right)\hat{\nabla}_{\alpha}\varphi &=& 0.\ \ \
\end{eqnarray}
\end{widetext}
Moving the above equation to $k$-space yields
\begin{widetext}
\begin{eqnarray}\label{eq:scalar_eom_k}
\left(K_{\sigma}+2\sigma K_{\sigma\sigma}\right)\xi'' + \left[\left(2K_\sigma-2\sigma K_{\sigma\sigma}-4\sigma^2K_{\sigma\sigma\sigma}\right)\frac{a'}{a} + \left(6\sigma K_{\sigma\sigma}+4\sigma^2K_{\sigma\sigma\sigma}\right)\frac{\varphi''}{\varphi'}\right]\xi'\nonumber\\
+ \left[K_{\sigma}k^2+\frac{{\rm d}^2\ln A(\varphi)}{{\rm d}\varphi^2}\left(\bar{\rho}^{(\nu)}-3\bar{P}^{(\nu)}\right)a^2+\frac{{\rm d}^2A(\varphi)}{{\rm d}\varphi^2}\left(\bar{\rho}^{(c)}+\bar{\rho}^{(b)}\right)a^2\right]\xi\nonumber\\
+ K_\sigma\varphi'k\mathcal{Z} + \frac{{\rm d}\ln A(\varphi)}{{\rm d}\varphi}\left(\bar{\rho}^{(\nu)}\Delta_{(\nu)}-3\mathcal{X}^{p,(\nu)}\right)a^2 + \frac{{\rm d}A(\varphi)}{{\rm d}\varphi}\left(\bar{\rho}^{(c)}\Delta_{(c)}+\bar{\rho}^{(b)}\Delta_{(b)}\right)a^2\nonumber\\
+ \left(K_{\sigma}+2\sigma K_{\sigma\sigma}\right)\varphi'w' + \left[2\left(K_{\sigma}+5\sigma K_{\sigma\sigma}+2\sigma^2K_{\sigma\sigma\sigma}\right)\varphi'' + \left(K_{\sigma}-4\sigma K_{\sigma\sigma}-4\sigma^2K_{\sigma\sigma\sigma}\right)\frac{a'}{a}\varphi'\right]w &=& 0.
\end{eqnarray}
\end{widetext}
{This equation is numerically solved in our modified {\sc Camb} code to compute the scalar field perturbations, the contribution of which, through Eqs.~(\ref{eq:sf_emt_com1} - \ref{eq:sf_emt_com4}), is included in the computation of curvature variables in Eqs.~(\ref{einsteink1} - \ref{einsteink6}).}

Using this equation, one could check that the total energy-momentum tensors for matter species that couple to the scalar field (cold dark matter, baryons and massive neutrinos) {satisfies Eq.~(\ref{non}) up to linear order and are not conserved (though recall that the energy momentum tensors for photons and massless neutrinos are not affected by this coupling and therefore are indeed individually conserved). However, the total energy momentum tensor, including the contribution from the scalar field, is conserved as can also be easily checked.} Such a check of the global conservation equations constitutes a robust validation of the equations derived so far.

\section{Numerical results}\label{sec:results}

\subsection{Numerical implementation and model parameters}

For the numerical implementation one needs to specify the following: (i) the functional form of $K(\sigma)$ in Eq.~(\ref{eq:scalar_lagrangian}) and its parameters, (ii) the functional form of $A(\varphi)$ in Eq.~(\ref{eq:conformal_transform}) and the parameters therein, and (iii) the value of the K-mouflage mass scale $M$, which appears in the definition $\sigma M^4=\frac{1}{2}\nabla^\mu\varphi\nabla_\mu\varphi$.

In this paper, we follow \cite{bv2014a,bv2014b} and consider
\begin{eqnarray}
K(\sigma) &\equiv& -1 + \sigma + K_0\sigma^m,
\label{K-def-K0-m}
\end{eqnarray}
in which $K_0$ is a real dimensionless parameter and $m\geq2$ a dimensionless integer. {Note that the model contains an effective cosmological constant ${\cal M}^4$ even though there is no explicit $\Lambda$ term in the actions Eqs.~(\ref{eq:action}, \ref{eq:S_m}).} We consider also
\begin{eqnarray}
A(\varphi) &\equiv& \exp\left(\beta\varphi/M_{\rm Pl}\right),
\end{eqnarray}
where $\beta$ is a another dimensionless model parameter that determines the strength of the coupling. Note that if $\beta = 0$, then $A(\varphi) = 1$, which corresponds to the standard uncoupled case.

For numerical considerations, it is more convenient to treat the scalar field with units of $M_{\rm Pl}$, i.e., we make the following field redefinition:

\begin{eqnarray}
\varphi &\leftarrow& \varphi/M_{\rm Pl}.
\end{eqnarray}
We can also write $M^4$ as
\begin{eqnarray}
M^4 &\equiv& \lambda^2M_{\rm Pl}^2H_0^2,
\end{eqnarray}
in which $\lambda$ is the newly-defined dimensionless parameter. The condition that the K-mouflage field drives the current accelerated cosmic expansion implies that $\lambda\sim\mathcal{O}(1)$.

According to these considerations, the K-mouflage model is specified by four dimensionless parameters -- $\left\{K_0, m, \beta, \lambda\right\}$. However, the value of $\lambda$ can be fixed by the condition that the K-mouflage field has a present-day energy density that makes the Universe spatially flat (non-flat models can also be considered, but these are beyond the scope of the current paper):
\begin{eqnarray}\label{eq:lambda_fixing}
\Omega_{\varphi0}\ =\ 1 - \Omega_{\gamma0} - \Omega_{c0} - \Omega_{b0} - \Omega_{r0} - \Omega_{\nu0} ,\ \ \ \
\end{eqnarray}
where $\Omega_{i0} \equiv \bar{\rho}_{i0}/\bar{\rho}_{\mathrm{cr}0}$ is the fractional energy density of the $i$-th species today {and $\bar{\rho}_{\mathrm{cr}0} = 3H_0^2M_{\rm Pl}^2$ is the critical density} {(note that, for generality, we have included both massless ($r$) and massive ($\nu$) neutrinos)}. This way the dimensionality of the K-mouflage part of the parameter space is reduced from four to three. The expression for $\sigma$ can then be written as
\begin{eqnarray}
\sigma &=& \frac{1}{2H_0^2\lambda^2}\nabla^\mu\varphi\nabla_\mu\varphi,
\end{eqnarray}
and the coupling function as
\begin{eqnarray}
A(\varphi) &\equiv& \exp\left(\beta\varphi\right).
\end{eqnarray}
Accordingly, in Eqs.~(\ref{eq:scalar_eom_background}, \ref{eq:scalar_eom_linear}, \ref{eq:scalar_eom_k}), $\varphi$ and $\xi$ are considered as dimensionless, provided the energy densities and pressure for dark matter, baryons and massive neutrinos are divided by $M_{\rm Pl}^2$. We will use the redefined equations in our numerical calculations, but for brevity will not present these redefined equations here.

For numerical convenience, linear perturbations are implemented in the frame where $w= 0$. This is the so-called cold dark matter frame (which is coincident with the synchronous gauge) in standard uncoupled models, but in the coupled cases the frame comoving with cold dark matter and the one where $w=0$ are not the same. This will not affect our conclusions, since the choice of frame has a negligible impact on the matter power spectrum on the scales where we have data, and since CMB temperature anisotropies and the lensing potential are frame-independent.

We set $\xi$ and $\xi'$ to zero as our initial conditions for the scalar field perturbation, and we have checked that changing them to values different from zero does not affect the numerical results noticeably, provided the values are not too exotic (non-exotic here means that the values chosen should guarantee that the K-mouflage density perturbation is much smaller than its background density to remain in the linear perturbation regime).

We choose to implement the homogeneous and linear perturbation dynamics of the K-mouflage model into the publicly available {\sc Camb} code \cite{camb}. We independently developed two versions of modified {\sc Camb} code which are in excellent agreement. We have also checked that our {\sc Camb} solutions satisfy the global conservation equations at both the background and linear perturbations levels, and for all matter species. {For $\lambda$, we adopt a simple bisection trial-and-error method to find its value with a $10^{-5}$ accuracy. }
Our results also agree very well with the numerical solver used in \cite{bv2014a,bv2014b}. These robust tests make us confident {about our codes and results.}

In this paper, we are interested in measuring the deviations induced by the K-mouflage coupling relative to the standard $\Lambda$CDM paradigm. For this reason, we shall compare these two model predictions for the fixed set of cosmological parameters
\begin{eqnarray}
&  & \left\{ T_{\mathrm {CMB}}, N_{\rm eff}, \hat \Omega_{c0} h^2, \hat \Omega_{b0} h^2, h, n_{\mathrm s}, 10^9 A_{\mathrm s}, \tau \right\} \nonumber \\
& = &  \left\{2.73, {3.046,} 0.118, 0.0221, 0.68, 0.964, 2.17, 0.864 \right\}, \nonumber \\
\end{eqnarray}
where $h\equiv H_0/(100{\rm km/s/Mpc})$ is the dimensionless present day Hubble expansion rate, $n_s$, $A_s$ are, respectively, the scalar spectral index and amplitude (at $k_{\rm pivot} = 0.05 \ \mathrm{Mpc}^{-1}$) of the power spectrum of the primordial scalar fluctuations, {$T_{\mathrm {CMB}}$ is the CMB monopole temperature today (in K)}, $\tau$ is the optical depth to reionisation and $N_{\rm eff}$ is the effective number of neutrino-like relativistic degrees of freedom.  {Note that $\hat \Omega_{b0}$ and $  \hat \Omega_{c0}$ are hatted and defined as $ \hat \Omega_{b,c} \equiv \hat \rho_{b,c} / \rho_{\mathrm{cr}}$. This choice is somewhat arbitrary but will have some impact when comparing K-mouflage effects on the matter and lensing potential power spectra.} {(The motivation for this choice is that it is the hatted matter density $\hat{\rho}$ that obeys the usual continuity equation and decreases as $a^{-3}$
at the background level.)}

\begin{figure}[h]
  \begin{center}
     \includegraphics[scale=0.65]{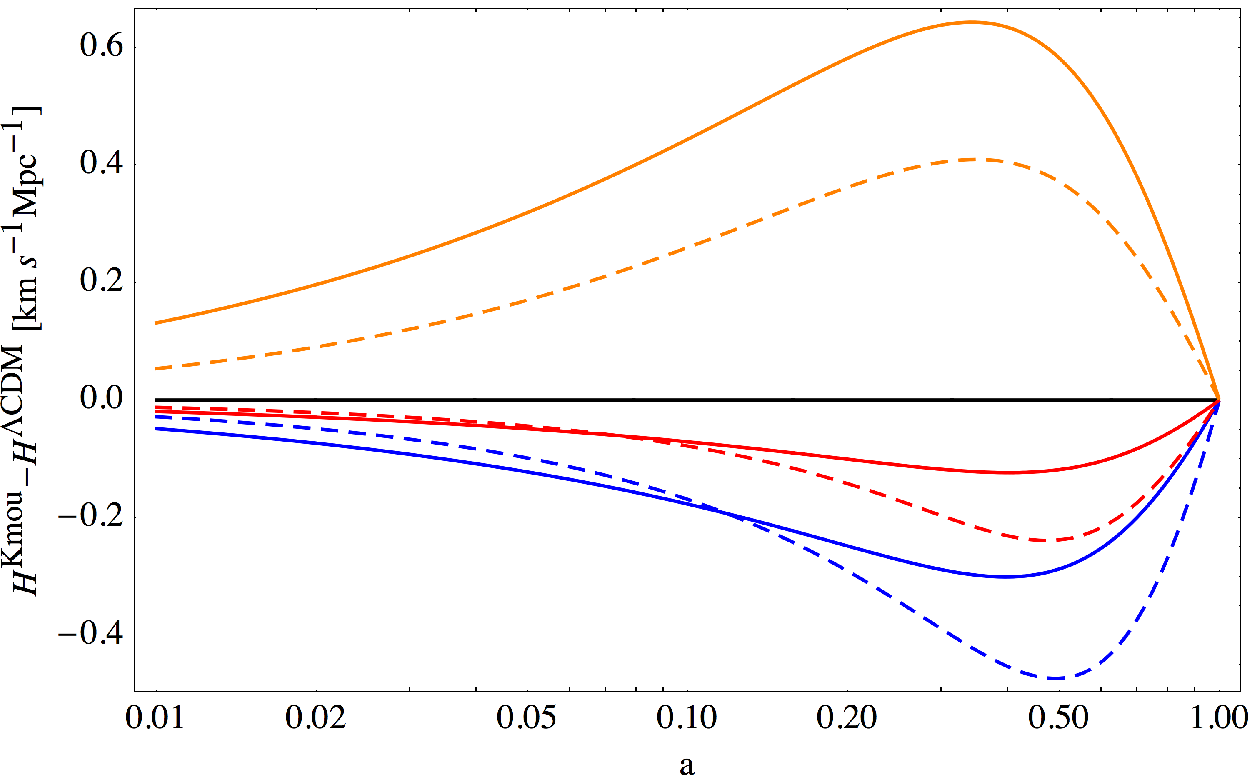}\\
      \includegraphics[scale=0.65]{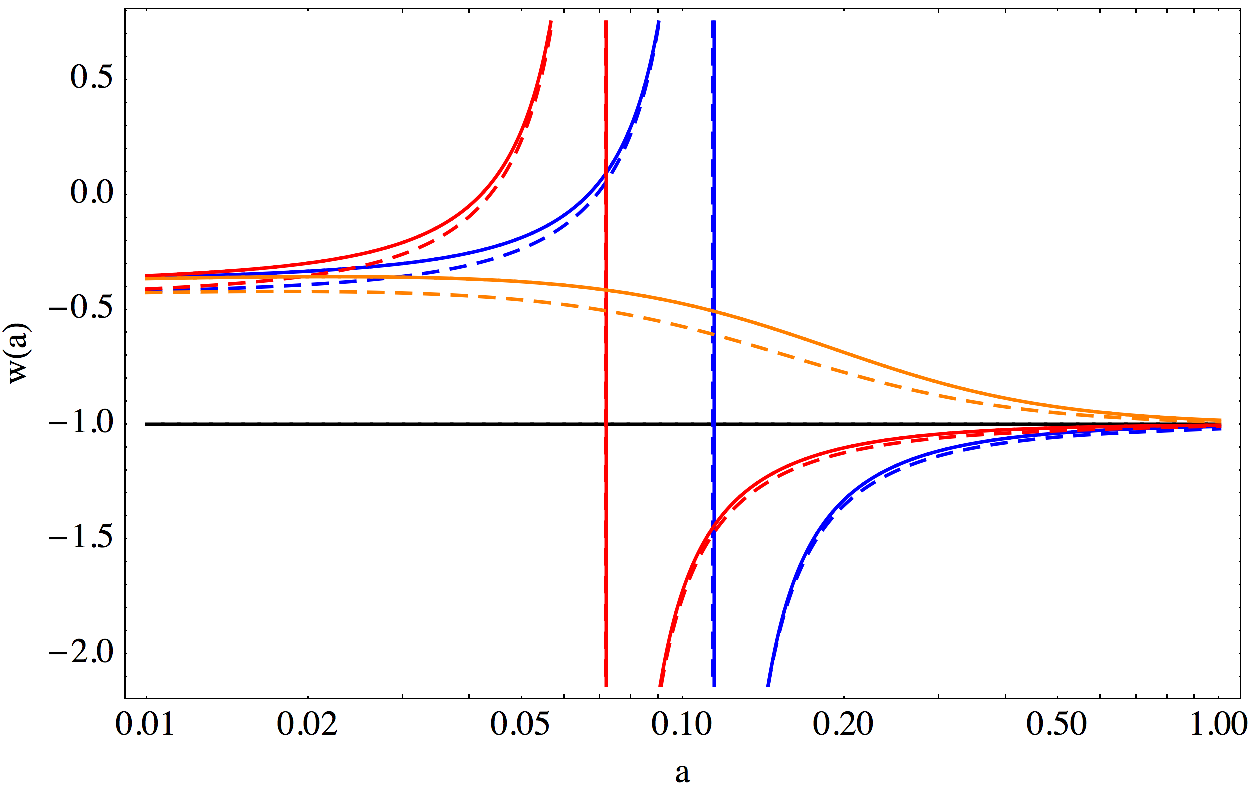}\\
     \includegraphics[scale=0.65]{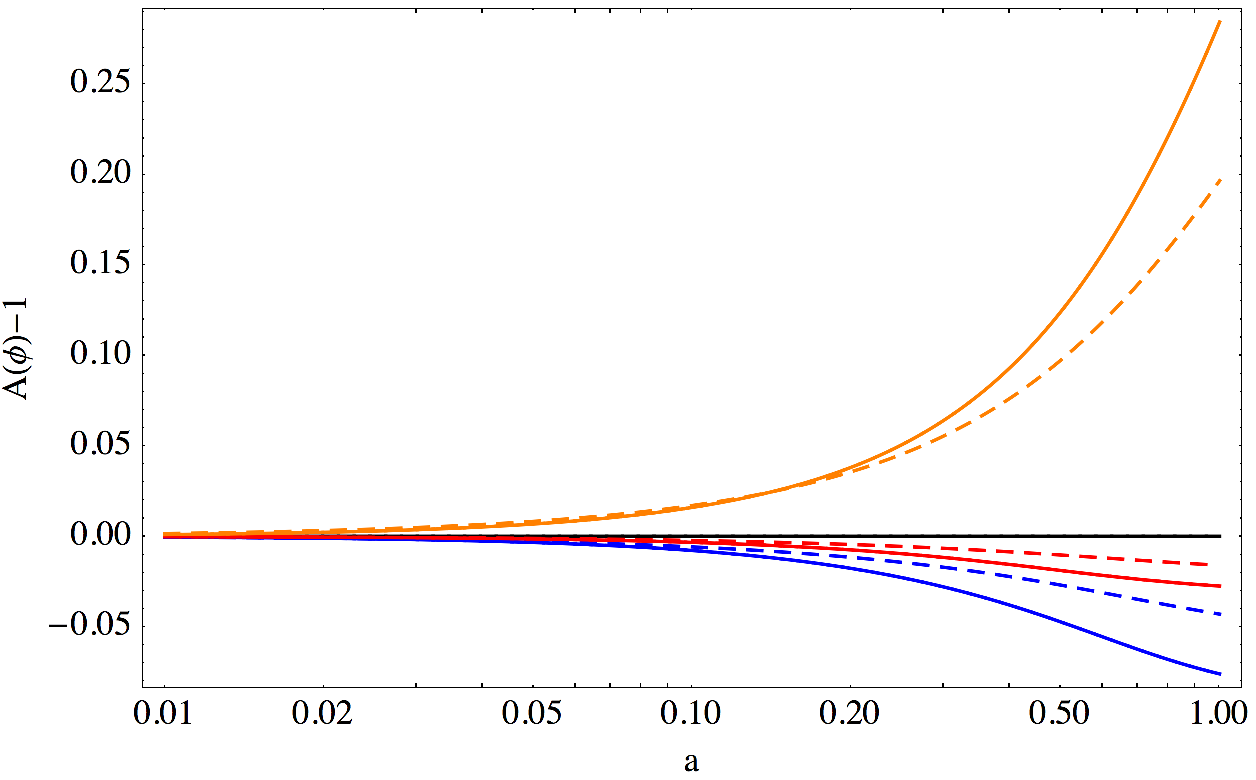}
  \end{center}
 \caption{Evolution with the scale factor of the difference between the K-mouflage and $\Lambda$-CDM Hubble rates, of the dark energy equation of state $w(a)$ and of $A(\varphi)-1$, for the following parameter set combinations:  $K_0 = 100$, $\beta = 0.2$ (blue), $K_0=100$, $\beta = 0.1$ (red), $K_0=-5$, $\beta=0.2$ (orange), for $m=3$ (solid lines) and $m=2$ (dashed lines).     \label{fig:bkgnd}}
\end{figure}

{In the following section, we will focus on the perturbation properties of our K-mouflage models.  In Fig.~\ref{fig:bkgnd}, we nevertheless have recalled how the most relevant background quantities evolve in K-mouflage cosmology.  The difference of the Hubble rate from $\Lambda$CDM, the equation of state, and the factor $A(\varphi)-1$ are displayed for the same parameters that will be considered in the following subsections.  The Hubble rate is lower than in $\Lambda$CDM at late times when $K_0>0$ and larger when $K_0<0$. Similarly, the equation of state at late times is either larger than -1 when $K_0<0$ or smaller than $-1$ when $K_0>0$.  Notice that in the latter case, it passes through a singularity at earlier times (but the product $w(a) \Omega_\varphi$ remains finite~\cite{bv2014a,bv2014b}).  Finally, we have plotted the factor $A(\varphi)$ which enters in the time evolution of particle masses such as electrons and quarks.  The mass of the proton, to leading order, is only dependent on the QCD scale which is independent of $\varphi$, as the K-mouflage field couples conformally to matter and leaves no effect on gauge fields. The factor $A(\varphi)$ also plays an important role in the perturbative analysis which follows. When $K_0>0$ it becomes lower than one, and inversely for $K_0<0$. }

Below, we analyse the impact the K-mouflage field has on the CMB temperature, CMB lensing potential and linear matter power spectra. We shall focus on a number of combinations of K-mouflage parameters to illustrate the relatively rich phenomenology of the model, paying particular attention to the degeneracies between the K-mouflage parameters and the summed mass of active neutrinos, $\Sigma m_\nu$. This will help us predict the types of observational constraints that can be placed upon this model.

For the time being, we analyse the following three scenarios: (i) K-mouflage with $\Sigma m_\nu = 0$, (ii) $\Lambda$CDM model with $\Sigma m_\nu \neq 0$, and (iii) K-mouflage with $\Sigma m_\nu \neq 0$. {Notice that
the effects of K-mouflage cannot easily be separated into the ones coming from a modification of the background cosmology compared to $\Lambda$CDM and a change in the perturbation evolution. Indeed, the effects of K-mouflage are driven by the presence of the coupling to matter $\beta$ and the non-linear terms in the Lagrangian. Setting $\beta\ne 0$ affects both the background cosmology and perturbations. If we were to analyse only the effect of $\beta$ on perturbations by keeping the background cosmology similar to $\Lambda$CDM (or vice versa), the equations of motion would not be consistent.}

\subsection{K-mouflage with $\Sigma m_\nu = 0$}\label{sect:Kwithoutnu}

\begin{figure*}[h]
  \begin{center}
   \includegraphics[scale=0.75]{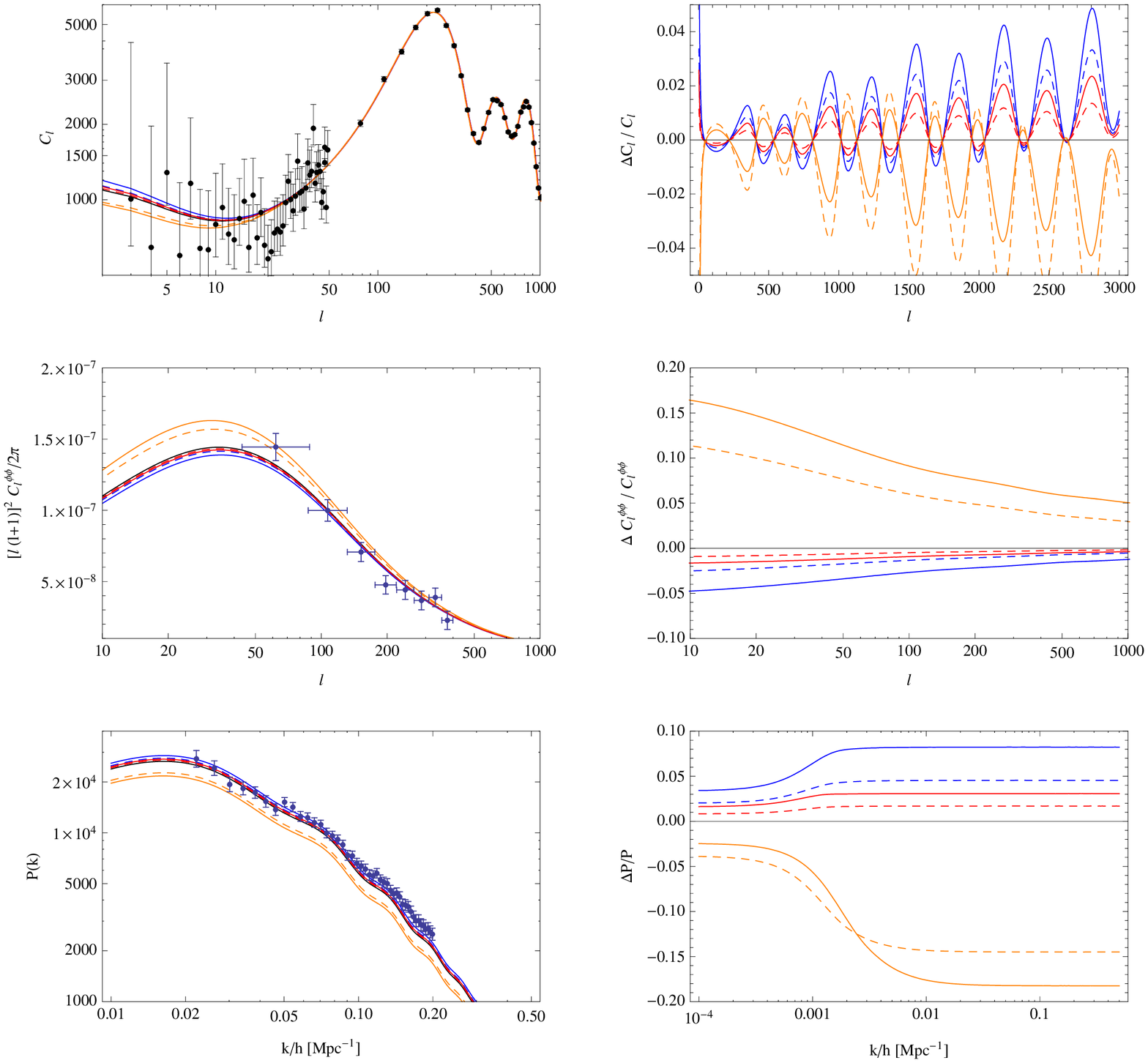}
  \end{center}
  \caption{CMB temperature angular power spectrum (top left), lensing potential power spectrum (middle left) and matter power spectrum (bottom left), as well as the corresponding relative difference between the K-mouflage and LCDM models (right panels), for the following parameter set combinations:  $K_0 = 100$, $\beta = 0.2$ (blue), $K_0=100$, $\beta = 0.1$ (red), $K_0=-5$, $\beta=0.2$ (orange), for $m=3$ (solid lines) and $m=2$ (dashed lines).  {The $\Lambda$CDM model used in the ratios is that with $\Sigma m_\nu = 0$ (solid black).} In the upper left and middle left panels, the data points with errorbars correspond, respectively, to the CMB temperature and lensing potential power spectra as measured by the Planck satellite \cite{planckxvi}. In the lower left panel, the data points show the SDSS-DR7 Luminous Red Galaxy (LRG) host halo power spectrum as presented in \cite{reid2010}.
   \label{fig:kmouflage}
 }
\end{figure*}

Considering first the effects on the matter power spectrum (bottom panels of Fig.~\ref{fig:kmouflage}), compared to the $\Lambda$CDM paradigm, we find that the result depends {qualitatively} on the sign of $K_0$. In particular, the K-mouflage model predicts more clustering than $\Lambda$CDM for $K_0 > 0$, and less clustering for $K_0 < 0$. On super horizon scales ($k \lesssim 2 \times 10^{-4} h$Mpc$^{-1}$), the modifications are scale-independent on the matter power spectrum.  On very large scales, terms involving powers of $k$ {become} negligible, which effectively eliminates the $k$-dependence from the equations. The modifications to $\Lambda$CDM on these large scales are driven by the modified expansion history, time variation of particle masses and clustering of the K-mouflage field (similar to that of the quintessence field on horizon scales).

For $k \gtrsim 0.01$ $h$Mpc$^{-1}$, the modifications are again scale-independent but the size is different. {This `plateau' in the relative difference from $\Lambda$CDM} is reached at smaller scales, if the deviation from $\Lambda$CDM is larger. On these sub-horizon scales, the terms involving powers of $k$ dominate in the equations, and the static approximation of \cite{bv2014b} holds, in which the scalar field density perturbation is negligible compared with the matter density perturbation (see {Section III.C1 of \cite{bv2014b}}). In this regime, the scalar field affects matter clustering through the modified expansion history, the varying particle masses, the fifth force, and the frictional force, which is itself a consequence of varying particle masses.

The bottom panels of Fig.~\ref{fig:kmouflage} also show that increasing the coupling strength $\beta$ leads to a stronger deviation from $\Lambda$CDM. This can be seen by comparing the blue ($\beta = 0.2$) and the red ($\beta = 0.1$) curves. Finally, for the K-mouflage models studied here, the effect of increasing the exponent $m$ is to boost the size of the modifications on all scales, if $K_0 > 0$. However, for $K_0 < 0$ (orange curves), we find that increasing $m$ increases the difference from $\Lambda$CDM on small scales, but suppresses it on large scales. {The detailed interplay of the impact of $K_0$, $m$ and $\beta$ on the growth of structure gives room for degenerate effects to arise. Some of these degeneracies might be broken by considerations of theoretical stability \cite{bv2014c} and/or observational constraints with different datasets (as we discuss below).}


The above results for the matter power spectrum are in good agreement with the estimations presented in \cite{bv2014b}. In the latter, it is shown that if $K_0 > 0$ is sufficiently large, then the K-mouflage model approaches $\Lambda$CDM. {We have confirmed this result with our {\sc Camb} versions, as well.} This suggests that the data from galaxy clustering (lower left panel of Fig.~\ref{fig:kmouflage}) {should} not put any upper limits on $K_0$, given that $\Lambda$CDM currently provides a reasonably good fit. Our results indicate that for $K_0 \sim \mathcal{O}(100)$ and $\beta \sim \mathcal{O}(0.1)$, the size of the deviation from $\Lambda$CDM is at the level of a few percent. We have also checked that decreasing $K_0$ and increasing $\beta$ boosts these differences further (not shown). From this we anticipate that current and future data should at least be able to place lower bounds on $K_0$ and upper bounds on $\beta$. The stringency of such bounds can only be fully determined through a detailed exploration of the parameter space. Nevertheless, a robust comparison between theory and galaxy clustering data requires also a proper modelling of the effects of galaxy and halo bias, redshift space distortions, and mode couplings on smaller scales induced by nonlinearities in the density field. All of these can only be properly addressed with N-body simulations, which is beyond the scope of the present work.

Due to these complications in comparing linear theory predictions with the large scale clustering of galaxies, it is likely that the CMB data (which is more robust and less prone to the effects of nonlinearities) will be more useful in constraining the K-mouflage model. The effects of the K-mouflage field on the CMB temperature power spectrum are shown in the top panels of Fig.~\ref{fig:kmouflage}. On small angular scales (high $\ell$'s), the relative difference from $\Lambda$CDM shows a series of oscillations that are roughly in phase opposition for the two values of $K_0$ shown. These oscillations of the relative difference follow from small horizontal shifts in the CMB power spectrum (barely visible in the upper left panel of Fig.~\ref{fig:kmouflage}) caused by the modifications to the expansion history in the K-mouflage model. The fact that these oscillations are in phase opposition for $K_0 = 100$ and $K_0 = -5$, indicates that these two cases shift the overall spectrum in opposite directions. Indeed, as first shown in \cite{bv2014a}, if $K_0 > 0$, then the Hubble expansion rate is smaller than in $\Lambda$CDM at late times. This shifts the spectrum towards higher $\ell$. Conversely, the spectrum gets shifted towards lower $\ell$ values if $K_0 < 0$. On large scales (low $\ell$), we find again that the deviations from $\Lambda$CDM depend qualitatively on the sign of $K_0$. This region of the CMB power spectrum is mostly determined by the integrated Sachs-Wolfe (ISW) effect, which is a secondary anisotropy induced on the temperature of CMB photons as they cross time-evolving gravitational potentials. The ISW effect is sensitive to the details of the late-time background expansion history, but in the K-mouflage models, the fifth force can also have a strong impact on the time variation of the potential. However, for these very large angular scales, the cosmic variance makes it difficult for stringent constraints to be derived. As for the case of the matter power spectrum, changes in the values of the coupling strength $\beta$ and exponent $m$ can amplify the size of the modifications to $\Lambda$CDM.

The K-mouflage model has also an important effect on the lensing potential power spectrum (middle panels of Fig.~\ref{fig:kmouflage}). For the range of $\ell$-values spanned by the Planck data, we find that the  {two} values  of $K_0$  lead to different amplitudes for the spectrum, indicating that these data may be able to put strong constraints on $K_0$. For the parameter values shown, the differences to $\Lambda$CDM are more pronounced at lower $\ell$ for which there is currently no data available. The amplitude of the CMB lensing potential power spectrum can also be affected by the values of $\beta$ and $m$, and as a result, we expect that current data may be able to place constraints on these parameters as well.

It is often said that both the matter power spectrum and the lensing potential power spectrum are sensitive probes of the clustering of matter in the Universe. The linear matter power spectrum is the Fourier transform of the two-point correlation function of the linear density {\it contrast} of matter, $\delta = \delta\rho_m/\bar{\rho}_m$. The lensing potential, on the other hand, is a weighted projection of the gravitational potential on the two-dimensional sky, which is obtained by integrating along the lines of sight from today up to the recombination epoch \cite{lc2006}. The CMB lensing potential power spectrum therefore probes the matter density {\it perturbations}, since the latter directly control the gravitational potential via Eq.~(\ref{einsteink2}). Hence, it might be confusing to observe that some of the K-mouflage parameters seem to have opposite effects on the amplitudes of these two spectra. In particular, the case for $K_0 = -5 < 0$ in Fig.~\ref{fig:kmouflage}, boosts the amplitude of the lensing potential power spectrum, but suppresses that of the matter power spectrum, and vice-versa for $K_0 = 100 > 0$. This seems contradictory since both probes are expected to be proportional to the amount by which matter clusters.

The above apparent tension follows from a nontrivial consequence of the scalar coupling in the K-mouflage model.
{For all the model considered, the input cosmological parameters $\hat \Omega_{b0} $ and $\hat \Omega_{c0} $ are identical and, as already mentioned, they are defined as the fraction of hatted matter densities, scaling as $a^{-3}$ and being conserved at the background and perturbed levels.   This choice is {somewhat} arbitrary and alternatively we could have normalized the models with fixed values of $ \Omega_{b0}$ and $  \Omega_{c0} $.   Background and linear perturbations in our modified {\sc Camb} are solved in the Einstein frame, where the hatted matter densities and {\it perturbations} (but not the density {\it contrasts}) are multiplied by $A(\varphi)$.   Indeed, the nonmininal coupling of nonrelativistic particles to the scalar field  induce a time variation of the particle mass in the Einstein frame and this extra scaling on the particle masses translates directly into the energy density of the matter particles, at both the background and perturbation levels (recall the discussion about Eq.~(\ref{eq:hat_T_munu}) ).   Since the gravitational potential is related to the {Einstein-frame} matter {\it perturbations}, it is also sensitive to the effects of the coupling.   On the other hand, the density contrast is the ratio of two densities, thus it is not affected by the coupling and is frame independent.   This explains the opposite effect we get on the matter and lensing potential power spectra:  considering negative values of $K_0$, the resulting fifth-force weakens the density contrast and the matter power spectrum, but on the other hand,  $A(\varphi) > 1$ at $z=0$ which amplifies the density perturbations sufficiently to change the sign of the deviation of the gravitational potential from $\Lambda$CDM in the Poisson equation.  The amplified gravitational potential then leads to an enhancement of the lensing potential power spectrum. A similar conclusion can be drawn when $K_0 > 0$.}
This rather nontrivial aspect of the K-mouflage model illustrates its rich phenomenology, and will be the focus of a more in-depth analysis in future work{, in which we will present a full analysis of the parameter space of K-mouflage models with $m=3$. This should help to disentangle the degeneracies between parameters, and will be useful to determine whether the discriminatory effects between lensing and matter power spectra can be used to further constrain the model while preserving  a good agreement with CMB measurements by the Planck mission.}

\subsection{$\Lambda$CDM with $\Sigma m_\nu \neq 0$}

\label{sect:lcdmwithnu}

\begin{figure*}[h]
  \begin{center}
     \includegraphics[scale=0.75]{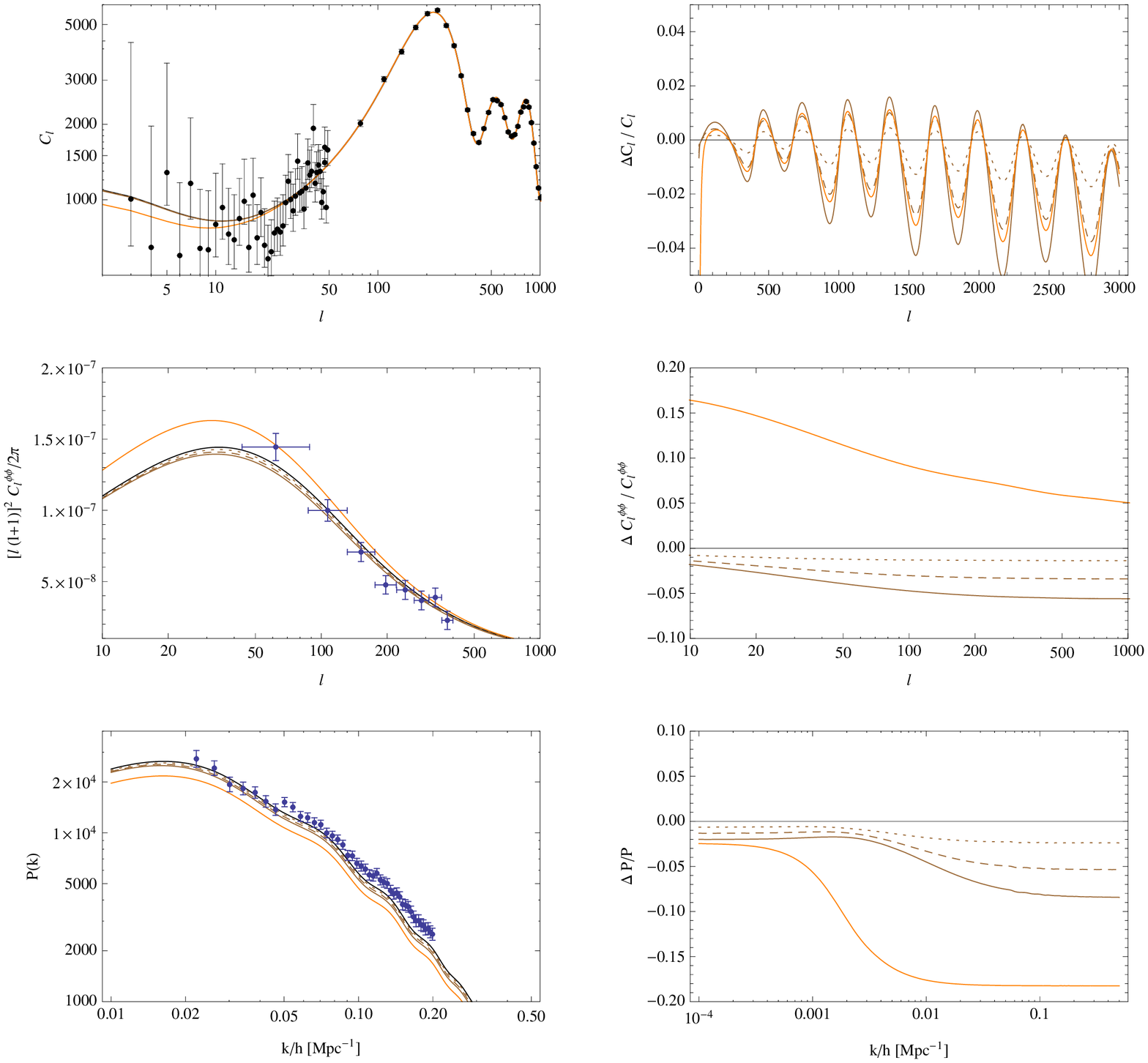}
  \end{center}
 \caption{CMB temperature angular power spectrum (top left), lensing potential power spectrum (middle left) and matter power spectrum (bottom left) for $\Lambda$CDM with $\Omega_\nu h^2 = 0.64 \times 10^{-3}$ (dotted brown) corresponding to $\sum_{\nu}m_{\nu} = 0.06$ eV,  $\Omega_\nu h^2 = 1.28 \times 10^{-3}$ (dashed brown) corresponding to $\sum_{\nu}m_{\nu} = 0.12$ eV and  $\Omega_\nu h^2 = 1.92 \times 10^{-3}$ (solid brown) corresponding to $\sum_{\nu}m_{\nu} = 0.18$ eV. We consider three active neutrinos with a degenerate mass spectrum. The results for the K-mouflage model with $K_0 = -5$, $\beta = 0.2$, $m=3$ are also displayed for comparison (solid orange line). The right panels show the corresponding relative differences to $\Lambda$CDM with $\Sigma m_\nu = 0$  {(solid black)}. In the upper left and middle left panels, the data points with errorbars correspond, respectively, to the CMB temperature and lensing potential power spectra as measured by the Planck satellite \cite{planckxvi}. In the lower left panel, the data points show the SDSS-DR7 Luminous Red Galaxy (LRG) host halo power spectrum as presented in \cite{reid2010}.    \label{fig:massivenu}
 }
\end{figure*}

Before analysing the effects of massive neutrinos in the K-mouflage model, it is instructive to remind ourselves of their role in standard $\Lambda$CDM. This is shown in Fig.~\ref{fig:massivenu} for three values of $\Sigma m_\nu$. We consider three active neutrinos with a degenerate mass spectrum, because for the current level of precision of the data one can safely ignore the mass splittings. For fixed $\Omega_{c0}h^2$ and $\Omega_{b0}h^2$, increasing the value of $\Sigma m_\nu$ increases the expansion rate at late times, after the neutrinos become nonrelativistic. Consequently, adding massive neutrinos also leads to the appearance of oscillations with $\ell$ when one takes the relative difference to a model without massive neutrinos (upper right panel of Fig.~\ref{fig:massivenu}). In particular, increasing $\Sigma m_\nu$ shifts the power spectrum slightly towards lower $\ell$, which is opposite to the effect of positive values of $K_0$ displayed in Fig.~\ref{fig:kmouflage} (note that the oscillations induced by $\Sigma m_\nu$ in the $\Lambda$CDM model are in phase opposition to those induced by $K_0$ > 0). This suggests that the peak positions of the CMB temperature data might determine a strong degeneracy between $K_0$ and $\Sigma m_\nu$. The impact neutrinos have on larger angular scales should be harder to distinguish because of the weaker constraining power of the CMB data there.

Furthermore, massive neutrinos also lower the amplitudes of the linear matter and CMB lensing potential power spectra.  {Note that here, contrary to the effects of the scalar coupling, the changes in the amplitude of these two spectra are consistent with one another.}
 The presence of a sufficiently large fraction of massive neutrinos can also lead to scale dependence in the growth of the matter fluctuations, because of the free streaming of massive neutrinos (cf.~lower panels of Fig.~\ref{fig:massivenu}).

\begin{figure*}[h]
 \begin{center}
    \includegraphics[scale=0.75]{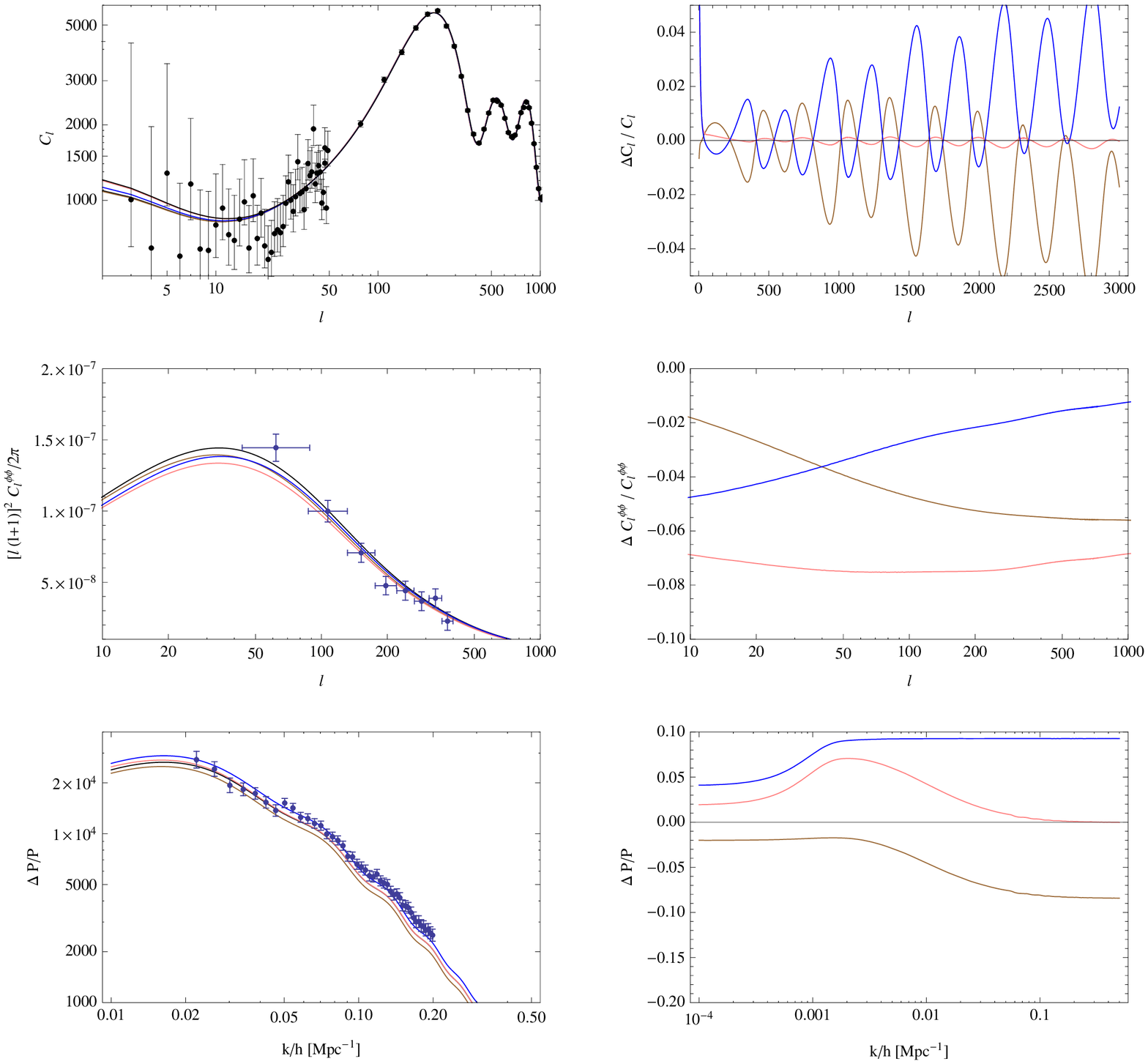}
  \end{center}
  \caption{CMB temperature angular power spectrum (top left), lensing potential power spectrum (middle left) and matter power spectrum (bottom left), as well as the corresponding relative difference between the K-mouflage and $\Lambda$CDM models (right panels), for $K_0 = 50$, $\beta = 0.2$, $m=3$ and $\Sigma m_\nu = 0.18\ {\rm eV}$ ($\Omega_\nu h^2 = 1.92 \times 10^{-3}$, in pink). The $\Lambda$CDM model used in the ratios is that with $\Sigma m_\nu = 0$ (solid black). For comparison, the results for the LCDM model with massive neutrinos and $\Omega_\nu h^2 = 1.92 \times 10^{-3}$ (brown) and for the K-mouflage model {with same parameters and $ \Sigma m_\nu = 0 $ (blue)} have been displayed. In the upper left and middle left panels, the data points with errorbars correspond, respectively, to the CMB temperature and lensing potential power spectra as measured by the Planck satellite \cite{planckxvi}. In the lower left panel, the data points show the SDSS-DR7 Luminous Red Galaxy (LRG) host halo power spectrum as presented in \cite{reid2010}.
  \label{fig:kmounu}
 }
\end{figure*}

\subsection{K-mouflage with $\Sigma m_\nu \neq 0$}\label{sect:Kwithnu}

Figure \ref{fig:kmounu} serves to confirm and illustrate some of the degeneracies between the K-mouflage parameters and $\Sigma m_\nu$, which have been anticipated in the discussion above. In particular, in terms of the high-$\ell$ part of the CMB temperature power spectrum in the K-mouflage models, we note that the presence of massive neutrinos can considerably cancel out the oscillations that appear in the relative difference to a $\Lambda$CDM model without massive neutrinos. Following from the discussion above, this is because the massive neutrinos and the K-mouflage parameters can shift the spectrum horizontally in opposite directions, and in such a way to preserve the position of the acoustic peaks (pink curve in the upper right panel of Fig.~\ref{fig:kmounu}).

Massive neutrinos cluster less strongly than cold dark matter, and so their presence leads to an overall suppression of the total matter clustering power. Their free streaming introduces also scale dependences in the growth of structure, which become more prominent on smaller scales. These effects could conspire with the scale-independent boosts in the clustering predicted by some of the K-mouflage parameter combinations on sub-horizon scales (e.g.~$K_0 > 0$) to leave the matter power spectrum nearly unchanged compared with $\Lambda$CDM (cf.~lower-right panel of Fig.~\ref{fig:kmounu} for $k \gtrsim 0.1 h/{\rm Mpc}$). However, the differences in the scale-dependent features introduced by $\Sigma m_\nu$ and the K-mouflage parameters on the growth of structure leave room for some breaking of degeneracies, although we recall, comparisons with galaxy clustering data require a better modelling of certain aspects of nonlinear structure formation.

Finally, since massive neutrinos also lower the amplitude of the lensing potential power spectrum, then some of the boosting effects of the K-mouflage field for negative values of $K_0$ (cf.~Fig.~\ref{fig:kmouflage}) can be cancelled out. However,  recall that the peak positions of the CMB are likely to determine a degeneracy between larger values of $\Sigma m_\nu$ and positive values $K_0$, not negative. As a result, combined constraints from the CMB temperature and lensing spectrum have the potential to partly break this degeneracy. For instance, the {middle panels} of Fig.~\ref{fig:kmounu} show that increasing $\Sigma m_\nu$ on the K-mouflage model with $K_0 > 0$ (pink curve), further suppresses the amplitude of the lensing power spectrum, compared to a $\Lambda$CDM model without massive neutrinos. Consequently, if the CMB peak positions could cope with large massive neutrino fractions for a positive $K_0$, this may still lead to an amplitude of the lensing power spectrum that is too low to be compatible with the observations.

\section{Summary and Discussion} \label{sect:conclusions}

\subsection{Summary}

In this paper, we have derived the fully covariant and gauge invariant linearly perturbed equations for cosmologies where a scalar degree of freedom couples directly to matter. In our derivation we have analysed, in detail, the case for each of the species that make up the energy content of the Universe. We aimed at being comprehensive, in the hope that the equations presented in this paper can serve as useful references for future works.

Although our equations are general, we have focused specifically on the case where the scalar field is a K-mouflage field. The Lagrangian structure of such a field is characterised by non-canonical kinetic terms that can hide the effects of the coupling to matter in regions where the gravitational acceleration (i.e.~first derivatives of the gravitational potential) exceeds some threshold. The study presented here, however, focuses on linear theory, for which the effects of the screening mechanism do not play a role. We have solved our set of equations in a suitably modified version of the {\sc Camb} code.

One of our main goals was to determine the impact of the K-mouflage model on observables such as the CMB temperature, CMB lensing potential, and linear matter power spectrum. We want to compare our results with those of the standard $\Lambda$CDM paradigm, and as a result, we have used a fixed set of cosmological parameters for both models. With this spirit, we only allowed ourselves to vary the summed mass of the three active neutrinos (to illustrate potential degeneracies) and the parameters that enter the K-mouflage Lagrangian.

We found that the coupled K-mouflage field modifies the background dynamics and hence shifts the CMB temperature power spectrum horizontally. This translates into a series of oscillations when we look at the relative difference to $\Lambda$CDM. For certain K-mouflage parameters, however, this effect can be cancelled by having massive neutrinos. This can potentially lead to interesting degeneracies between the modification to gravity and neutrinos masses. The K-mouflage model can also have a visible impact on the larger angular scales of the CMB temperature power spectrum, through its modifications to the ISW effect. However, it is unlikely that this signal would lead to significant constraints, given the large size of the error bars due to cosmic variance.

Our results show that matter clustering can also be significantly affected by the coupled K-mouflage field, especially on sub-horizon scales, where the scalar coupling with matter has the strongest impact. Again, massive neutrinos can cancel out some of the effects, but introduce also scale-dependences on the growth that might be used to break some degeneracies and impose constraints on the model parameters. We remark that a proper use of galaxy clustering data to constrain models of modified gravity should only be performed after a more careful analysis of the nonlinear regime of structure formation (see e.g.~Sec.~IV.D.~of~\cite{barreira2014a}).

The K-mouflage models which enhance (suppress) the amplitude of the linear matter power spectrum, seem to suppress (enhance) the amplitude of the lensing potential power spectrum. This seems to be contradictory at first sight, since both observables should probe the overall matter clustering. This is because of a rather nontrivial effect of the scalar coupling on the magnitude of the gravitational potentials. The latter are essentially determined by the sizes of the (absolute) density perturbations, which are rescaled in the same way as matter particle masses. However, when one computes the density contrast to calculate the matter power spectrum, the time dependences in the matter density perturbation and background matter density caused by varying particle masses cancel out. This effectively leads to different qualitative predictions on the amplitudes of the matter and lensing potential power spectra. Such a feature of the K-mouflage model may lead to interesting constraints on the model's parameter space.

{Although we have presented our results in the case of cubic K-mouflage models (with $m=3$), they are quite general as long as the K-mouflage Lagrangian is dominated by a monomial of power $m>1$ with coupling constant $K_0$. In this case, the models which have both an early Universe behaviour with a vanishing influence of the scalar field in the radiation era and a static screening of the scalar interaction must satisfy $K_0>0$ for odd $m$ and $K_0<0$ for even $m$. The latter leads to ghost instabilities and should be discarded. In the case of odd $m$'s, when $K_0>0$, all the conclusions stand similarly to the cubic case, in particular structure growth is enhanced while the lensing spectrum is reduced. Moreover, the excursion of the field $\varphi$ is small enough compared to the Planck scale that one can always expand the function $A(\varphi)$ to linear order with no significant difference. }

\subsection{Discussion and Outlook}

We conclude by briefly comparing the predictions of the K-mouflage model with those of other recently studied modified gravity models.

The background evolution of the covariant Galileon model does not admit a $\Lambda$CDM limit (see e.g.~\cite{barreira2012}). In particular, the effective dark energy equation-of-state parameter of the Galileon field is phantom (smaller than $-1$) in the recent past, which is similar to the K-mouflage model for $K_0 > 0$ \cite{bv2014a}. This lowers the expansion history at late times, which shifts the CMB temperature power spectrum to higher $\ell$. In the case of the Galileon model, this feature is responsible for a strong preference of the model for significantly large massive neutrino fractions ($\Sigma m_\nu \gtrsim 0.4$ at $\sim 6\sigma$) \cite{barreira2014b, barreira2014c}. We have seen in Sec.~\ref{sec:results} that a similar trend might also arise in the K-mouflage model. However, the nontrivial effects that a scalar coupling can have on the lensing power spectrum (cf.~Sec.~\ref{sect:Kwithoutnu}) are absent in the Galileon model \cite{galileon2}. Nonlocal formulations of gravity \cite{mm2014, dirian2014, barreira2014d} can also lead to background solutions that are different from $\Lambda$CDM, in a way that its impact on the peak positions might also prefer nonzero values for $\Sigma m_\nu$. On the other hand, chameleon models typically possess free functions that can be tuned to yield exact $\Lambda$CDM expansion histories.

On linear sub-horizon scales, the effects of the K-mouflage and Galileon fields on the growth of structure are both scale independent. On even smaller scales, however, the two models react differently to the nonlinear density field due to their different screening mechanisms. For Galileons, N-body simulations \cite{barreira2013b, li2014quartic, barreira2014a} show that the effects of the screening mechanism start to become important on scales $k \gtrsim 0.1 h/{\rm Mpc}$, which correspond to the typical size of dark matter clusters. In the K-mouflage model, on the other hand, the screening mechanism only becomes important on much smaller scales $k \gtrsim 10 h/{\rm Mpc}$ \cite{bv2014b, bv2014c}. On scales of $0.1~h{\rm Mpc}^{-1} \lesssim k \lesssim 10 h{\rm Mpc}^{-1}$, the nonlinear regime of structure formation in the K-mouflage model should therefore resemble more the case of Nonlocal gravity models \cite{barreira2014d}, for which the modifications to gravity are not screened. For chameleon models, the environmentally dependent Compton wavelength of the chameleon field leads to a scale-dependent growth, even on linear scales. However, the current observational constraints on chameleon model parameters essentially make these models' large scale structure predictions to be nearly undistinguishable from $\Lambda$CDM (see e.g.~\cite{lombriser2014}).

As we hope to have shown above, the variety and size of the observational signatures that characterise the K-mouflage model leave us with an interesting playground to explore several degeneracies with a number of cosmological parameters. In this paper, we have focused specifically on the case of massive neutrinos, but we note that degeneracies with other parameters, such as $h$ and $\Omega_{c0}h^2$, could also be present, because both have an impact on the expansion history and clustering strength. The K-mouflage model is different from other popular modified gravity models, making it of interest for further investigations. In particular, in a coming work we will determine its overall goodness-of-fit, by exploring the global cosmological parameter space with Monte Carlo Markov Chain methods.

\

\begin{acknowledgments}

We thank Antony Lewis for helpful discussions on the numerical implementation of massive neutrinos in the {\sc Camb} code. AB acknowledges support by FCT-Portugal through grant SFRH/BD/75791/2011. The work of SC is supported by the \textit{mandat de retour} program of the Belgian Science Policy (BELSPO). BL is supported by the Royal Astronomical Society and Durham University.  P.B.
acknowledges partial support from the European Union FP7 ITN
INVISIBLES (Marie Curie Actions, PITN- GA-2011- 289442) and from the Agence Nationale de la Recherche under contract ANR 2010
BLANC 0413 01.

\end{acknowledgments}

\end{document}